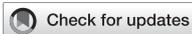

# Proton electromagnetic generalized polarizabilities

N. Sparveris*

Department of Physics, Temple University, Philadelphia, PA, United States

Electromagnetic polarizabilities are fundamental properties of the proton that characterize its response to an external electromagnetic (EM) field. The generalization of the EM polarizabilities to non-zero four-momentum transfer opens up a powerful path to study the internal structure of the proton. They map out the spatial distribution of the polarization densities in the proton, provide access to key dynamical mechanisms that contribute to the electric and magnetic polarizability effects, and allow for the determination of fundamental characteristics of the system, such as the electric and magnetic polarizability radii. This article reviews our knowledge about proton EM generalized polarizabilities (GPs). An introduction is given to the basic concepts and the theoretical framework, which is then followed by a discussion that emphasizes the recent developments and findings of the virtual Compton scattering (VCS) experiments and future perspectives on the topic.

KEYWORDS

proton, polarizabilities, hadrons, virtual Compton scattering, nucleon

## 1 Introduction

Explaining how the proton emerges from the interactions of its quark and gluon constituents is a central goal of modern nuclear physics. In order to understand the underlying dynamics of quarks and gluons in the proton as governed by quantum chromodynamics (QCD), the theory of the strong interaction and theoretical calculations require experimental guidance and confrontation with precise measurements of the fundamental properties of the system. Starting with the simple electromagnetic (EM) process, $\gamma^* p \rightarrow p$, one can access the charge and magnetization densities in the proton via the measurement of the proton elastic form factors. A wealth of information on the dynamics of the proton is hidden in the response of the charge and magnetization densities of the proton to an external electromagnetic field, i.e., in the EM polarizabilities. The classical concept of polarizabilities was extended to the case of the nucleon with the first Compton scattering measurements on the proton in the 1950s [1–5]. A polarizability is a fundamental structural constant for any composite system, and the two scalar polarizabilities of the proton [6, 7] – the electric, $\alpha_E$, and the magnetic, $\beta_M$ – can be interpreted as the response of the proton structure to an external electric and magnetic field, respectively. They describe how easily the charge and magnetization distributions inside the proton are distorted by the EM field and provide the net result of the spatial distributions of the system. In order to measure the polarizabilities, one must generate an electric ($\vec{E}$) and a magnetic ($\vec{H}$) field. The electric field strength needed to induce any appreciable polarizability of the nucleon can be estimated as the ratio of the average energy level spacing in the nucleon to the size of the nucleon, i.e., $\approx 100\,\text{MeV}/(e\,\text{fm}) = 10^{23}$ Volt/m. Static electric field strengths of this intensity are not available in a laboratory. Nevertheless, in the case of the proton, this is provided by the photons in the Compton scattering process. A classical estimate of the electric field strength of a 100 MeV photon in Compton scattering from the nucleon is approximately $10^{23}$ V/m, allowing the Compton scattering process to be used for the experimental access of the nucleon polarizabilities.

The two scalar polarizabilities appear as second-order terms in the expansion of the real Compton scattering (RCS) amplitude in the energy of the photon, as shown in Eq. 1.

$$H_{eff}^{(2)} = -4\pi\left(\frac{1}{2}\alpha_E \vec{E}^2 + \frac{1}{2}\beta_M \vec{H}^2\right). \quad (1)$$

Moving on to higher orders, the third-order term depends on the nucleon spin, and the corresponding polarizabilities are called *spin polarizabilities* [8]. Here, there is no analog in classical electrodynamics, but they practically describe the coupling of the induced EM moments with the nucleon spin, and unlike the scalar polarizabilities, they are not invoked by static EM fields. Coming back to the two scalar polarizabilities, one can provide a simplistic description of these quantities based on the resulting effect of an electromagnetic perturbation applied to the nucleon constituents. An electric field moves the positive and negative charges inside the proton in opposite directions. The induced electric dipole moment is proportional to the electric field, and the proportionality coefficient is the electric polarizability, which quantifies the stiffness of the proton. On the other hand, a magnetic field has a different effect on the quarks and pion cloud within the nucleon, giving rise to two different contributions in the magnetic polarizability, a paramagnetic and a diamagnetic contribution, respectively. Compared to the atomic polarizabilities, which are of the size of the atomic volume, the proton electric polarizability $\alpha_E$ (see Figure 1) is much smaller than the volume scale of a nucleon [6], and the typical units adopted for the polarizabilities are $10^{-4}$ fm$^3$. The small magnitude underlines the stiffness of the proton, a direct consequence of the strong binding of its constituents, and indicates the intrinsic relativistic character of the system.

The first Compton scattering measurement of the proton polarizabilities using a tagged photon beam was reported by [9]. More Compton scattering measurements on the proton followed in the next few years, e.g., by the LEGS group [10] and the TAPS at the Mainz Microtron (MAMI) setup [11] in 2001. Recent advances in proton polarizability measurements involve the use of linear polarization as an analyzer to measure the electric and magnetic polarizabilities independently from each other, not relying on the Baldin sum rule (see Eq. 2) [12, 13] (i.e., not relying on the value of their sum), where the sum rule expresses the sum of dipole polarizabilities in terms of an integral of the total photoabsorption cross-section $\sigma_T$:

$$\alpha_E + \beta_M = \frac{1}{2\pi^2}\int_{\nu_0}^{\infty} \nu \frac{\sigma_T(\nu)}{\nu^2}. \quad (2)$$

Two such measurements were reported recently. At the Mainz Microtron, Compton scattering measurements on the proton below the pion threshold were performed using a tagged photon beam and a nearly $4\pi$ detector [14]. The electron beam, with an energy of 883 MeV, impinged on a 10-μm-thin diamond radiator, producing a linear polarized photon beam via coherent Bremsstrahlung. The resulting photon beam passed a 3-mm-diameter lead collimator and was incident on a 10-cm-long liquid hydrogen target, while the final state particles were detected using the Crystal Ball/TAPS setup [14]. The two static polarizabilities of the proton were determined as $\alpha_E = (10.99 \pm 0.16 \pm 0.47 \pm 0.17 \pm 0.34)~10^{-4}$ fm$^3$ and $\beta_M = (3.14 \pm 0.21 \pm 0.24 \pm 0.20 \pm 0.35)~10^{-4}$ fm$^3$. The second experiment [15] was performed at the High Intensity Gamma-Ray Source (HIGS) facility at the Triangle Universities Nuclear Laboratory, reporting for two polarizabilities $\alpha_E = (13.8 \pm 1.2 \pm 0.1 \pm 0.3)~10^{-4}$ fm$^3$ and $\beta_M = (0.2 \pm 1.2 \pm 0.1 \pm 0.3)~10^{-4}$ fm$^3$. The tensions between the MAMI and HIGS measurements illustrate the difficulty in conducting these experiments and highlight the need to be cautious in the treatment of the experimental uncertainties. The first simultaneous extraction of the six leading-order proton polarizabilities, namely, the two scalar and the four spin-dependent polarizabilities, was recently performed using an innovative bootstrap-based fitting method and the complete set of experimental world data [16], leading to $\alpha_E = (12.7 \pm 0.8_{(fit)} \pm 0.1_{(model)})~10^{-4}$ fm$^3$ and $\beta_M = (2.4 \pm 0.6_{(fit)} \pm 0.1_{(model)})~10^{-4}$ fm$^3$.

The generalization of the two scalar polarizabilities in four-momentum transfer space [17], $\alpha_E(Q^2)$ and $\beta_M(Q^2)$, is an extension of the static electric and magnetic polarizabilities obtained in RCS. The generalized polarizabilities (GPs) are studied through measurements of the virtual Compton scattering (VCS) process [17]: $\gamma^*p \to p\gamma$. VCS is accessed experimentally through the ep $\to$ ep$\gamma$ reaction, where the incident real photon of the RCS process is replaced by a virtual photon. The virtuality of the incident photon ($Q^2$) sets the scale of the observation and allows one to map out the spatial distribution of the polarization densities in the proton, while the outgoing real photon provides the EM perturbation to the system. The meaning of the GPs is analogous to that of the nucleon form factors. Their Fourier transform will map out the spatial distribution density of the polarization induced by an EM field. They probe the quark substructure of the nucleon, offering a unique insight into the underlying nucleon dynamics. They allow for the determination of fundamental characteristics of the system, such as the EM polarizability radii, and frequently enter as input parameters in various scientific problems, e.g., in the hadronic two-photon exchange corrections, which are needed for the precise extraction of the proton charge radius from muonic hydrogen spectroscopy measurements [18]. The following sections review our knowledge about proton scalar generalized polarizabilities. First, the theoretical context and theoretical calculations of the GPs are discussed, followed by a review of the virtual Compton scattering experiments with an emphasis on recent developments, ongoing projects, and the consideration of future prospects. This paper adds to and complements previous reviews including those by [17, 19, 20] on the VCS; [18, 21] on dispersion relations for Compton scattering; [22] on effective field theory and Compton scattering experiments; [23] on the pion, nucleon, and kaon polarizabilities; and [24] on covariant baryon chiral perturbation theories (BChPTs), while we can refer to [25] for the extension of the discussion to the spin structure of the nucleon that is not covered here.

## 2 Theoretical framework for VCS

### 2.1 VCS at low energy and the LET

The VCS [17] reaction can be obtained from RCS by replacing the incident real photon with a virtual photon $\gamma^*$. Experimentally, VCS can be accessed as a subprocess of the reaction $e(k) + N(p) \to e$ $(k') + N(p') + \gamma(q')$. The particle four-momentum vectors are denoted as $k^\mu$ and $k'^\mu$ for the incoming and scattered electrons, $q^\mu$ and $q'^\mu$ for the virtual photon and final real photon, and $p^\mu$ and $p'^\mu$ for the initial and final protons, respectively. The modulus of the three momenta is denoted as $q = |\mathbf{q}|$. The variables that are indexed "cm" are in the center-of-mass (c.m.) frame of the initial proton + virtual photon, i.e., the c.m. of the Compton process. The kinematics of the reaction are defined by five independent variables. The set of variables that is typically adopted involves $(q_{cm}, q', \epsilon, \theta_{cm}, \phi)$, where $\epsilon$ (see Eq. 3) is the virtual photon polarization parameter, i.e.,

$$\epsilon = 1 \bigg/ \left[1 + 2 \frac{q_{lab}^2}{Q^2} \tan^2(\theta'_{e\,lab}/2)\right]. \tag{3}$$

$q_{cm}$ and $q'$ are the three-momentum moduli of the virtual and final photons in the c.m., respectively, and $\theta_{cm}$ and $\phi$ are the c.m. angles of the Compton process, i.e., the polar and azimuthal angles of the outgoing real photon with respect to the virtual photon. In the above reaction, the real final photon can be emitted by either the electron or nucleon. The first contribution corresponds to the Bethe–Heitler (BH) process, which is well known and calculable from QED with the nucleon electromagnetic form factors as an input. The second part involves the VCS subprocess. VCS can, in turn, be decomposed further into a Born term, where the intermediate state is a nucleon, and a non-Born term, which contains all nucleon excitations and meson-loop contributions, as shown in Figure 2. The non-Born amplitude $T^{NB}$ contains the physics of interest and is parametrized at low energy by the nucleon GPs. The three amplitudes add up coherently (see Eq. 4) to form the total photon electroproduction amplitude, i.e.,

$$T^{ep \to ep\gamma} = T^{BH} + T^{Born} + T^{NB} = T^{BH} + T^{VCS}. \tag{4}$$

In obtaining the non-Born amplitude, a multipole expansion [26] is performed in the c.m. frame, yielding the multipoles $H_{NB}^{(\rho'L',\rho L)S}(q'_{cm}, q_{cm})$. Here, $L$ ($L'$) represents the angular momentum of the initial (final) electromagnetic transition in the $(\gamma^* p \to \gamma p)$ process, and $S$ differentiates between the spin-flip ($S = 1$) or non-spin-flip ($S = 0$) transition at the nucleon side. [$\rho$ ($\rho'$) = 0, 1, 2] characterizes the longitudinal ($L$), electric ($E$), or magnetic ($M$) nature of the initial (final) photon. The GPs are obtained as the limit of these multipoles when $q'_{cm}$ tends to be zero at arbitrarily fixed $q_{cm}$. At this strict threshold, the final photon has zero frequency, its electric and magnetic fields are constant, corresponding to a "static field," and the GPs represent the generalization at finite $q_{cm}$ of the polarizability in classical electromagnetism. For small values of $q'_{cm}$, one may use the dipole approximation ($L' = 1$), corresponding to electric and magnetic final-state radiation that is dipolar only. In this case, angular momentum and parity conservation lead to 10 different dipole GPs [26] that the nucleon crossing combined with charge conjugation symmetry then reduces further to 6 independent GPs [27, 28], namely, the 2 scalar GPs ($S = 0$) and the 4 spin-dependent (or spin-flip) GPs ($S = 1$). The two scalar GPs, the electric and magnetic, are thus defined as

$$\alpha_E(Q^2) = -\frac{e^2}{4\pi} \cdot \sqrt{\frac{3}{2}} \cdot P^{(L1,L1)0}(Q^2), \tag{5}$$

$$\beta_M(Q^2) = -\frac{e^2}{4\pi} \cdot \sqrt{\frac{3}{8}} \cdot P^{(M1,M1)0}(Q^2), \tag{6}$$

with $e^2/4\pi = \alpha_{QED} = 1/137$, and at $Q^2 = 0$, they coincide with RCS static electromagnetic polarizabilities $\alpha_E$ and $\beta_M$.

At low energy $q'$ of the emitted photons, the low-energy theorem (LET) and the low-energy expansion (LEX) for VCS [26] state that the non-Born term starts at order $q'$, whereas the Born term enters at $1/q'$. They offer a model-independent approach that allows us to express the VCS cross-section in terms of the GPs and to access GPs through experiments via the $ep \to ep\gamma$ reaction. The LEX formula yields for the photon electroproduction cross-section below the pion-production threshold:

$$d\sigma = d\sigma_{BH+Born} + \Phi q'_{cm} \Psi_0 + O(q'^2_{cm}), \tag{7}$$

$$\Psi_0 = V_1 (P_{LL} - P_{TT}/\epsilon) + V_2 P_{LT}, \tag{8}$$

where, $d\sigma_{BH+Born}$ is the BH + Born cross-section, which is entirely calculable in QED and requires the nucleon elastic form factors $G_E$ and $G_M$ as inputs. It contains no polarizability effect and serves as an important cross-section of reference throughout the formalism. $(\Phi q'_{cm})$, $V_1$, and $V_2$ are kinematical factors. The term $(\Phi q'_{cm} \Psi_0)$ is where the GPs first appear in the expansion. $\Psi_0$ is the first-order polarizability term obtained from the interference between the BH + Born and non-Born amplitudes at the lowest order. It is therefore of order $q'^0_{c.m.}$, i.e., independent of $q'_{cm}$. In the phase-space factor $(\Phi q'_{cm})$, the term $q'_{cm}$ has been explicitly factored out in order to emphasize the fact that when $q'_{cm}$ tends to be zero, $(\Phi q'_{cm} \Psi_0)$ tends to be zero and the whole cross-section tends to be $d\sigma_{BH+Born}$. The $O(q'^2_{cm})$ term represents all the higher-order terms of the expansion and contains GPs of all orders. Below the pion-production threshold, $d\sigma_{BH+Born}$ dominates the cross-section, $\Psi_0$ is the leading polarizability term, and the higher-order terms $O(q'^2_{cm})$ are expected to be negligible. The $\Psi_0$ term contains three VCS response functions or structural functions, namely, $P_{LL}$, $P_{LT}$, and $P_{TT}$, as defined in Eqs 9, 10, 11, which are combinations of the lowest-order GPs.

$$P_{LL}(Q^2) = -2\sqrt{6} M_N \cdot G_E^p(Q^2) \cdot P^{(L1,L1)0}(Q^2), \tag{9}$$

$$P_{TT}(Q^2) = -3 G_M^p(Q^2) \cdot \frac{q_{cm}^2}{\tilde{q}^0} \cdot \big( P^{(M1,M1)1}(Q^2)$$
$$- \sqrt{2}\, \tilde{q}^0 \cdot P^{(L1,M2)1}(Q^2) \big), \tag{10}$$

$$P_{LT}(Q^2) = \sqrt{\frac{3}{2}} \cdot \frac{M_N \cdot q_{cm}}{\tilde{Q}} \cdot G_E^p(Q^2) \cdot P^{(M1,M1)0}(Q^2)$$
$$+ \left[\frac{3}{2} \frac{q_{cm} \sqrt{Q^2}}{\tilde{q}^0} \cdot G_M^p(Q^2) \cdot P^{(L1,L1)1}(Q^2)\right], \tag{11}$$

where $M_N$ is the mass of the nucleon. Here, we can point out that $P_{LL}$ is proportional to the electric GP, $P_{LT}$ has a spin-independent part that is proportional to the magnetic GP, plus a spin-dependent part $P_{LT\,spin}$, and $P_{TT}$ is a combination of two spin GPs.

## 2.2 Dispersion relation formalism

The LEX of the VCS observables provides a model-independent method to analyze VCS experiments below the pion-production threshold in terms of structural functions that contain information about GPs. Nevertheless, the sensitivity of the VCS cross-section to the GPs is enhanced in the region between the pion-production threshold and the $\Delta$-resonance region, where the LEX does not hold.

Here, the dispersive approach provides a valuable theoretical framework to extract the GPs. In the DR formalism, the non-Born contribution to the VCS scattering amplitude is parametrized in terms of 12 independent amplitudes $F_i(Q^2, \nu, t)$, $i = 1, \ldots, 12$, which depend on 3 kinematical invariants, $Q^2$, $t$, and the crossing-symmetric variable $\nu = (s - u)/(4M_N)$ [29]. The GPs are expressed in terms of the non-Born part $F_i^{NB}$ at the points $t = -Q^2$ and $\nu = 0$. Assuming an appropriate analytic and high-energy behavior, these amplitudes fulfill unsubtracted DRs in the variable $\nu$ at fixed $t$ and fixed $Q^2$:

$$\operatorname{Re} F_i^{NB}(Q^2, \nu, t) = F_i^{pole}(Q^2, \nu, t) - F_i^B(Q^2, \nu, t) + \frac{2}{\pi} \mathcal{P} \int_{\nu_{thr}}^{+\infty} d\nu' \frac{\nu' \operatorname{Im} F_i(Q^2, \nu', t)}{\nu'^2 - \nu^2}, \quad (12)$$

where $F_i^B$ is the Born contribution as defined by [20, 26], whereas $F_i^{pole}$ denotes the nucleon pole contributions. Furthermore, Im $F_i$ are the discontinuities across the $s$-channel cuts, starting at the pion-production threshold $\nu_{thr} = m_\pi + (m_\pi^2 + t/2 + Q^2/2)/(2M)$. The validity of the unsubtracted DRs in Eq. 12 relies on the assumption that at high energies ($\nu \to \infty$, fixed $t$, and fixed $Q^2$), the amplitudes decrease fast enough such that the integrals converge. The high-energy behavior of the amplitudes $F_i$ was investigated [29, 30], with the finding that the integrals diverge for $F_1$ and $F_5$. For energies up to the $\Delta$-resonance region, one can saturate the $s$-channel dispersion integral by the $\pi N$ contribution, setting the upper limit of integration to $\nu_{\max} = 1.5$ GeV. The remainder can be estimated by energy-independent functions, which parametrize the asymptotic contribution due to $t$-channel poles, as well as the residual dispersive contributions beyond the value $\nu_{\max} = 1.5$ GeV. The asymptotic contribution to $F_5$ is saturated by the $\pi^0$ pole [29]. The asymptotic contribution to $F_1$ can be described phenomenologically as the exchange of an effective $\sigma$ meson in the same spirit as for unsubtracted DRs in the RCS case. The $Q^2$ dependence of this term is unknown. It can be parametrized in terms of a function directly related to the magnetic dipole GP $\beta_M(Q^2)$ and fitted to VCS observables. Furthermore, it was found that the unsubtracted DR for the amplitude $F_2$ is not so well saturated by $\pi N$ intermediate states only. The additional $s$-channel contributions beyond the $\pi N$ states can effectively be accounted for with an energy-independent function at fixed $Q^2$ and $t = -Q^2$. This amounts to introducing an additional fit function, which is directly related to the electric dipole GP $\alpha_E(Q^2)$. In order to provide predictions for VCS observables, it is convenient to adopt the following parametrizations for the fit functions:

$$\alpha_E(Q^2) - \alpha_E^{\pi N}(Q^2) = \frac{\alpha_E^{exp} - \alpha_E^{\pi N}}{\left(1 + Q^2/\Lambda_\alpha^2\right)^2}, \quad \beta_M(Q^2) - \beta_M^{\pi N}(Q^2)$$
$$= \frac{\beta_M^{exp} - \beta_M^{\pi N}}{\left(1 + Q^2/\Lambda_\beta^2\right)^2}, \quad (13)$$

where $\alpha_E$ and $\beta_M$ are the RCS polarizabilities, with superscripts $exp$ and $\pi N$ indicating the experimental value and the $\pi N$ contribution evaluated from unsubtracted DRs, respectively. In Eq. 13, the mass scale parameters $\Lambda_\alpha$ and $\Lambda_\beta$ are free parameters, not necessarily constant with $Q^2$, which can be adjusted by a fit to the experimental cross-sections.

The LEX and DR formalism have been widely tested in a series of experiments discussed in the following sections. Data below the pion-production threshold have been analyzed and compared with both the LEX and DR. Furthermore, the polarizabilities have been extracted via VCS measurements above the pion threshold by utilizing the DR formalism, and the results are in excellent agreement with those from the measurements below the pion-production threshold. A fundamental difference between the two (LEX vs. DR) analysis methods is that the DR formalism allows for the direct extraction of the scalar GPs by fitting the parameters $\Lambda_\alpha$ and $\Lambda_\beta$ to the data, whereas the LEX analysis provides access to structural functions depending linearly on both scalar and spin GPs [26]. In order to disentangle the scalar GPs, the contribution from the spin GPs to the structural functions has to be subtracted using a model, and for that, one typically uses the DR prediction for the spin GPs. The complementarity and synergy of the two frameworks extend further to the use of DR formalism toward the estimation of higher-order terms in the LEX analysis. The LEX and DR frameworks have so far illustrated remarkable agreement and consistency in their extracted values of the scalar GPs from all the experimental measurements.

## 2.3 Theoretical calculations of the generalized polarizabilities

Theoretical calculations for the virtual Compton scattering have been performed within a spectrum of frameworks that allow a complementary insight into the mechanisms that govern the GPs. The early calculations were conducted within the context of the non-relativistic constituent quark model (NRCQM) [26, 31, 32], which was later extended [33] to include relativistic effects by considering a Lorentz covariant multipole expansion of the VCS amplitude and a light-front relativistic calculation of the nucleon excitations. The constituent quark model calculations have some limitations since truncations are introduced that, e.g., lead to a violation of gauge invariance. Nevertheless, the calculations have been constructive in providing a first approximate estimate for the nucleon resonance contributions to the GPs. The resonance contribution to the GPs has been accounted for more accurately within the effective Lagrangian model (ELM). These calculations are founded on a fully relativistic effective Lagrangian framework, which includes baryon resonance contributions and $\pi^0$ and $\sigma$ exchanges in the $t$-channel [34, 35] or on calculations that use a coupled-channel unitary approach [36].

Calculations within the framework of the linear sigma model (LSM) and chiral effective field theories offer a different perspective through the prism of chiral symmetry and the pionic degrees of freedom. The LSM allows for a simplistic description of the nucleon that is based on relevant symmetries like Lorentz, gauge, and chiral invariance. These calculations [37, 38] have revealed the existence of relations between the VCS multipoles in addition to the usual constraints of parity and angular momentum conservation [27, 39]. The contribution of the pion-cloud effects can be accounted for using chiral perturbation theories (ChPTs), an expansion in the external momenta, and the pion mass ("$p$"-expansion). The VCS amplitude is consistent with electromagnetic gauge invariance, the pattern of chiral symmetry breaking in QCD, and Lorentz covariance to a given order of the small parameter $p \equiv \{P, m_\pi\}/\Lambda$, where $P$ stands for each component of the four momenta of the photons and the three momenta of the nucleons and $\Lambda$ is the breakdown scale of the theory. The first calculation [40, 41] was performed at $O(p^3)$ with only nucleons and pions as explicit degrees of freedom, using the heavy-baryon (HB) expansion for the nucleon propagators, which amounts to making an expansion in $1/M_N$ along with the expansion in $p$. The calculations were later extended to $O(p^4)$ and $O(p^5)$ [42, 43]. The inclusion of $\Delta(1232)$ as an explicit degree of freedom in the calculation of the VCS process was first accounted for by [44] by introducing the excitation energy of $\Delta(1232)$ as an additional expansion parameter ("$\epsilon$ expansion"). An alternative counting has also been proposed by [45] ("$\delta$-expansion"), and it was employed for the VCS process by [46] using a manifestly Lorentz invariant version of the baryon chiral perturbation theory (BChPT).

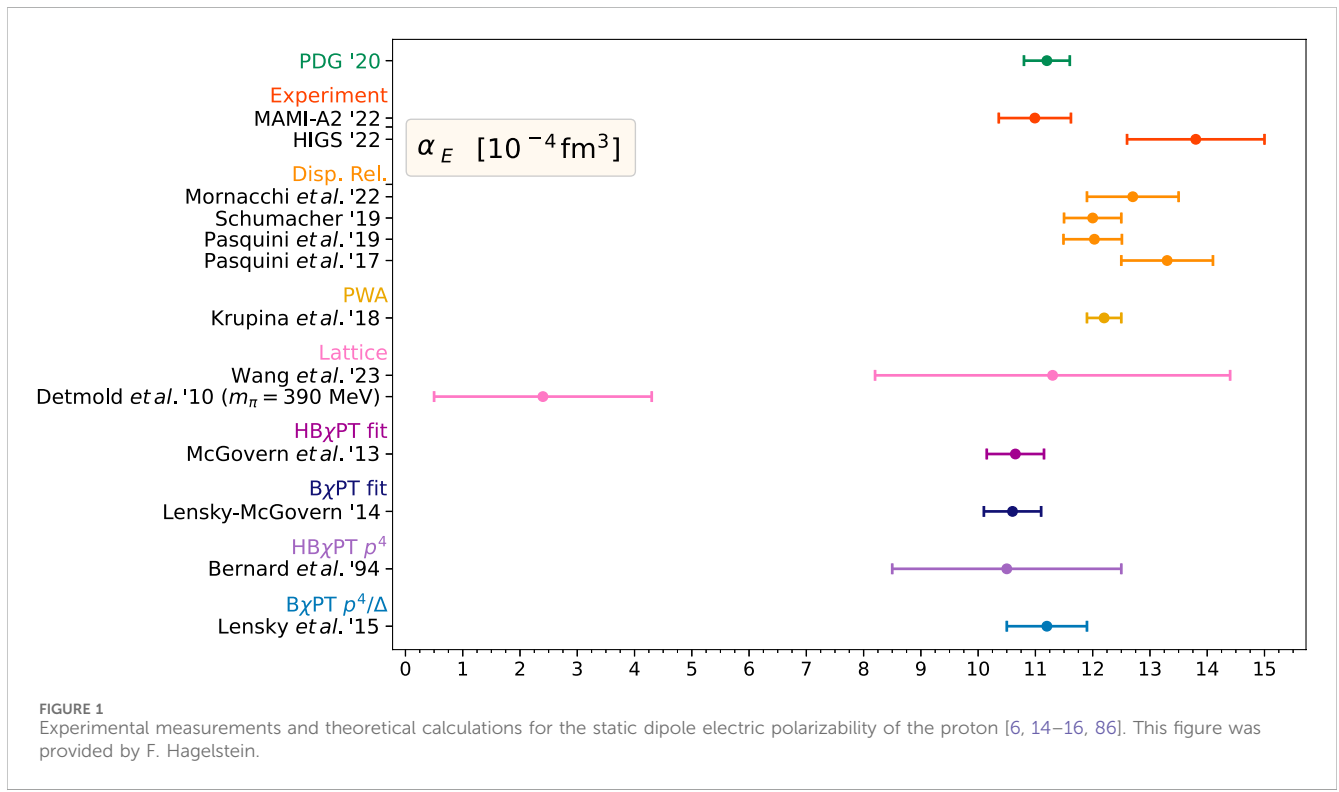

FIGURE 1
Experimental measurements and theoretical calculations for the static dipole electric polarizability of the proton [6, 14–16, 86]. This figure was provided by F. Hagelstein.

A representative group of the above calculations is shown in Figure 3. The differences in the $Q^2$ dependence of the calculations are driven by a number of parameters, i.e., in the NRCQM, the excited states of the nucleon are given by resonances, and the $Q^2$ behavior of the GPs is determined by the electromagnetic transition form factors. The LSM accounts for the excitation spectrum as pion–nucleon scattering states with a different $Q^2$ dependence and predicts a rapid variation at small momentum transfer and a smooth variation at higher momentum, while the ELM allows for both resonant and non-resonant contributions. Nevertheless, all these calculations underestimate the static electric polarizability at the real photon point ($Q^2 = 0$) by 30%–40%. For the magnetic GP, the pion cloud contributes to a positive slope at the origin, while the $N \to \Delta$ transition form factor drives the paramagnetic contribution, which decreases as a function of $Q^2$. The LSM describes only the negative diamagnetic component; the NRCQM accounts only for the positive paramagnetic contribution; and the interplay of the two competing effects can be observed in the ELM calculation. The calculation at the next-to-leading order in BChPT [46] is also shown in Figure 3. Here, the consideration of the pion-cloud effects allows us to overcome the shortcomings of the previous models in accounting for the magnitude of the static electric polarizability. Nevertheless, the BChPT prediction comes with a sizable theoretical uncertainty of ~30%, highlighting the need for the next order calculation.

## 2.4 Radiative corrections to virtual Compton scattering

The radiative corrections to virtual Compton scattering have been developed in analogy with the radiative corrections to elastic scattering. These corrections have been extensively studied by [47] and extended to the high-energy limit for deeply virtual Compton scattering by [48]. The main elements of the radiative corrections to the VCS experiments are summarized here, while a detailed description of the steps applied in the data analysis of the VCS experiments can be found in several works, such as those by [49–52].

The calculation distinguishes between virtual corrections that involve the exchange of a supplementary virtual photon and real radiative corrections that involve the emission of a supplementary real photon. The supplementary virtual photon is produced by a variety of processes, such as vacuum polarization, electron self-energy, or an additional loop between the lines of the reaction diagram. The correction term, $\delta_V$, varies slowly with $Q^2$ and is nearly constant as a function of the other variables. For the part of the real photon that can be emitted before or after the hadronic part of the interaction, the process can be expressed as the sum of two terms, $\delta_{R1}$ and $\delta_{R2}$. The first term depends on the maximum energy that the soft photon can reach and on the missing-mass squared cut that is used for each event in the data analysis, while $\delta_{R2}$ contains correction terms that do not involve such a dependence. The acceptance-dependent part is implemented in the simulation of the experiment, while the acceptance-independent correction is treated analytically and has a weak dependence on $Q^2$ and is nearly constant, excluding the proximity of the BH peaks. Real external corrections, $\delta'$, i.e., real radiation coming from another nucleus in the target, are treated in a classical way, as described by [53]. Thus, for the radiatively corrected cross-section, one obtains the following expression of Eqs 14, 15 [47]:

$$d\sigma_{exp}^{corr} = d\sigma_{exp}^{raw} \cdot F_{rad}, \quad (14)$$

with

$$F_{rad} = \left[1 + \delta_V + \delta_{R1} + \delta_{R2} + \delta'\right]^{-1}, \quad (15)$$

where $d\sigma_{exp}^{raw}$ is the raw measured cross-section. Beyond these primary corrections, one can improve further by adding smaller terms, e.g., the two-photon exchange contribution, $\delta_1$, and radiative corrections at the proton side, $\delta_2^{(0)}$ [47]. An example for $Q^2 = 0.3$ $GeV^2$ is at the kinematics of the first VCS experiment at MAMI and the recent experiment at JLab (VCS-I), $F_{rad} \simeq 0.95$ and $\simeq 0.99$, respectively. The $F_{rad}$ factor is determined with an accuracy of ±2%. Nevertheless, in the case of the azimuthal asymmetry measurements, the effect of these uncertainties in the extraction of the polarizabilities is significantly suppressed.

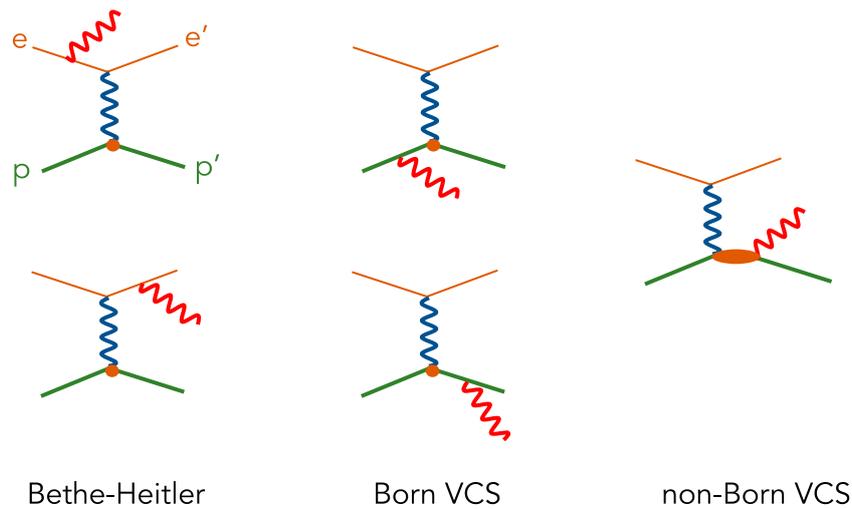

FIGURE 2
Feynman diagrams of photon electroproduction. The small circles represent the interaction vertex of a virtual photon with a proton considered a point-like particle, while the ellipse denotes the non-Born VCS amplitude.

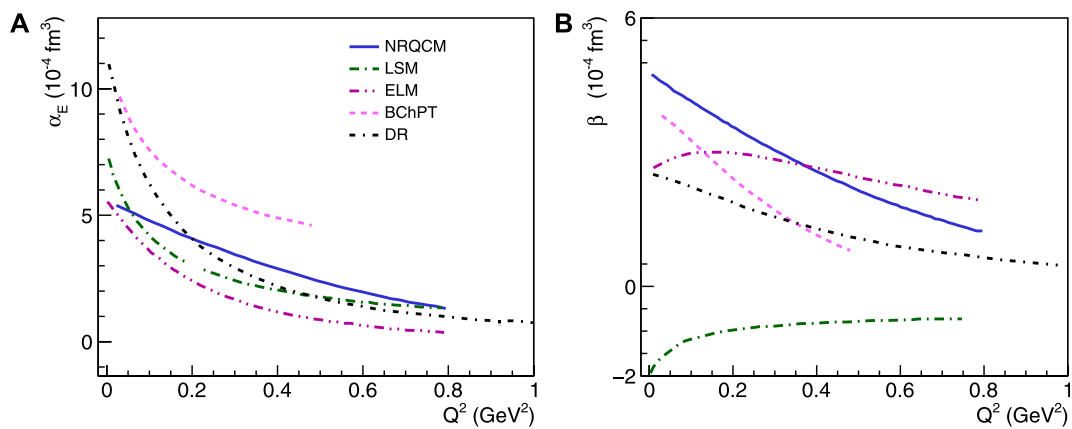

FIGURE 3
The theoretical predictions of BChPT [46], NRQCM [32], LSM [37], ELM [36], and DR [21, 29, 30] are shown for the electric generalized polarizability (A) and the magnetic generalized polarizability (B).

## 3 Experimental measurements of the generalized polarizabilities

### 3.1 Unpolarized cross-section measurements

The VCS experimental program brings together measurements that were conducted at the MAMI, MIT Bates, and JLab electron accelerator facilities. The measurements span a period of three decades, with the first one taking place in 1995 [54, 55]. The experiments followed a similar technique, i.e., utilized an electron beam, a liquid hydrogen target, and the detection of the two charged particles $e'$ and $p'$ in magnetic spectrometers. The exclusive reaction is then identified using the missing mass square of the undetected photon. The extraction of GPs from the VCS cross-section measurements is not direct and requires a fit made within a theoretical framework. The experiments use two different frameworks, with different domains of validity in $W$, that were discussed in the previous section: a model-independent one based on the LEX and a model-dependent one based on dispersion relations.

The LEX formalism is valid only below the pion-production threshold. The VCS amplitude is considered real, a property that holds only for $W < (M_N + m_\pi)$. As soon as hadronic intermediate states beyond that of the nucleon can be created on-shell, starting with a nucleon plus a pion, the VCS amplitude acquires an imaginary part, and the LEX formalism [26] is not valid anymore. The LEX fit is performed at fixed $q_{cm}$ and $\epsilon$ and yields the two structural functions $P_{LL} - P_{TT}/\epsilon$ and $P_{LT}$. The scalar GPs can then be deduced if one subtracts from these two structural functions their spin-dependent part, namely, $P_{TT}$ and $P_{LTspin}$. In the absence of available measurements of the spin GPs, this subtraction relies on a model calculation. In practice, $d\sigma_{BH+Born}$ is the cross-section without the polarizability effect, and the structural functions $P_{LL} - P_{TT}/\epsilon$ and $P_{LT}$ are obtained by fitting the deviation of $d\sigma_{exp}$ from $d\sigma_{BH+Born}$. The dispersion relations provide a powerful formalism to analyze VCS experiments both below and above the pion-production threshold. The imaginary part of the VCS amplitude is a central ingredient of the model, entering dispersive integrals saturated by $\pi N$ intermediate states. The existence of free parameters in the model related to the unconstrained part of $\alpha_E(Q^2)$ and $\beta_M(Q^2)$ allows us to perform an experimental fit in order to extract the scalar GPs. In practice, the DR fit is based on the comparison of the measured cross-section to the one calculated using the DR model and is less straightforward than the LEX fit since the structural functions or GPs do not appear in a simple analytic form in the model cross-section. Here, the spin GPs are fully constrained in the DR model and cannot be fitted. The formalism is suited for values of $W$ up to ~1.3 GeV, i.e., slightly above the $\pi\pi N$ threshold and covering most of the $\Delta(1232)$ resonance region.

The first experimental measurements involved the MAMI-I experiment [54, 55], which made use of the A1 experimental setup at MAMI (see Figure 4). The experiment was pioneering since it was the first to establish and implement all the experimental considerations of the VCS measurements, from the design of the experiment to the numerous aspects of the data analysis, e.g., the radiative corrections, dedicated Monte Carlo simulations, the implementation of the LEX fit methods for the GP extraction. The measurements targeted the region below the pion-production threshold, covering an extended range of photon energies, but were limited to in-plane angles. The first experiment at JLab [56, 57] explored the highest photon virtualities so far, in the range $Q^2 = 1\text{--}2\ GeV^2$. The GPs were found to be very small, pointing to a rapid fall-off with $Q^2$. For a common $Q^2$, data were taken both below and above the pion-production threshold, and the GPs were extracted independently for the two datasets. This allowed us to show for the first time that the GP extractions both below and above the pion-production threshold, using LEX and DR formalisms, respectively, yielded consistent results. It was also shown that the sensitivity to polarizabilities is amplified when one measures above the pion threshold and into the nucleon resonance region. The MIT Bates experiment [58, 59] was the first to use an extracted CW beam from the MIT Bates South Hall Ring, and it exploited the potential of the out-of-plane kinematics. It took advantage of the unique strengths of the out-of-plane spectrometer (OOPS) system to access large out-of-plane angles, combined with the ability to perform simultaneous measurements at different azimuthal angles, using up to four identical, lightweight magnetic spectrometers. This, in turn, allowed a significant reduction in the systematic uncertainties of the measurement. Data were taken at the smallest $Q^2$ so far (i.e., at 0.057 GeV$^2$), enabling the first extraction of the mean-square electric polarizability radius of the proton. This experiment was also the first to evidence a bias in the LEX fit and highlight the limitations of using this method of extraction for the GPs.

The above group of experiments took place during the first decade of the VCS program and shaped a preliminary understanding of the proton electric and magnetic GPs, covering a broad range in momentum transfer. The measurements came with relatively large experimental uncertainties, primarily for the magnetic polarizability, highlighting the experimental challenges in extracting the weak magnetic polarizability signal. One puzzling observation of the reported measurements involved the electric GP. Here, the data pointed to an enhancement of the $\alpha_E$ magnitude at $Q^2 = 0.33$ GeV$^2$ compared to the other measurements, which violated the theoretical expectation for a monotonic fall-off as a function of $Q^2$. Consequently, a follow-up experiment was conducted at MAMI, at the same $Q^2 = 0.33$ GeV$^2$, aiming to cross-check and confirm the findings of MAMI-I. The measurements of the MAMI-IV experiment [60] used the same experimental setup, focused on angular kinematics very similar to MAMI-I, and confirmed the results that were previously found.

The confirmation of the puzzling structure of $\alpha_E(Q^2)$ by two independent experiments motivated the need for further measurements. The experiments that followed focused on providing a finer $Q^2$ mapping of the polarizabilities and improving the precision of the results. The MAMI-V experiment [61] determined the electric GP at $Q^2 = 0.20$ GeV$^2$ from the cross-section and asymmetry measurements in the $\Delta(1232)$ resonance. Here, the $\alpha_E$ extraction was performed exclusively within the DR framework since the LEX formalism is not applicable in the $\Delta(1232)$ region. In parallel to the polarizability measurement, the experiment offered the first extraction of the $N \rightarrow \Delta$ quadrupole amplitude through the photon excitation channel, thus providing a stringent control to the model uncertainties of the $N \rightarrow \Delta$ transition form factors that have relied exclusively on measurements of the dominant pion–electroproduction channel. The MAMI-VI experiment [62] was performed at three values of $Q^2 = 0.10, 0.20,$ and 0.45 GeV$^2$. The structural functions and scalar GPs were extracted with high precision from cross-section data below the pion-production threshold, utilizing both LEX and the DR formalism. Out-of-plane kinematics were selected at each $Q^2$, in line with the approach of the MIT Bates experiment, but they covered a larger angular phase space. Capitalizing on the experience and lessons learned from the previous measurements, a novel bin selection method was developed, aiming to suppress the higher-order terms of the LEX. The DR model was utilized to provide an estimate of the higher-order terms of the LEX expansion. The theoretical cross-sections, $d\sigma_{DR}$ and $d\sigma_{LEX}$, were calculated using the same input values of structural functions. Since $d\sigma_{DR}$ includes all orders in $q'_{c.m.}$, the difference ($d\sigma_{DR} - d\sigma_{LEX}$) is a measure of the higher-order terms $O(q'^2_{c.m.})$ of Eq. 7, as given by the DR model. Accordingly, the following dimensionless estimator of Eq. 16 was constructed:

$$O(q'^2_{c.m.})_{DR} = \frac{d\sigma_{DR} - d\sigma_{LEX}}{d\sigma_{BH+Born}}, \quad (16)$$

at each point in the VCS phase space. The model-dependent estimator was used first in the design of the experiment in order to define kinematics, where $O(q'^2_{c.m.})$ is expected to be small. It was later employed in the data analysis of the experiment to study the behavior of the LEX fit under varying conditions. LEX fits were performed including a varying number of experimental bins, corresponding to gradually increased values of the $O(q'^2_{c.m.})_{DR}$ estimator by setting the condition $O(q'^2_{c.m.})_{DR} \leq K$ and letting the threshold $K$ vary. Polarizability LEX fits were performed, and their results were compared with and without the novel bin selection method, while a comparison was also done with the results of the DR fit analysis that is not bound to similar restrictions in the phase-space selection. The analysis concluded with the selection of $K_{optimal} = 0.025$ as the value providing the most reliable LEX fit. The studies indicated that the effect of the higher-order contributions becomes small and less of a concern at momentum transfers higher than 0.4 $GeV^2$. Careful consideration is to be given in the data analysis at lower-momentum transfers, where the resulting phase-space masking takes a typical shape, as shown in Figure 5. The refinements in the data analysis procedure of MAMI-VI allowed us to eliminate any bias arising from the higher-order contributions and minimize the systematic uncertainties in the extraction of the GPs. The refined analysis procedure of MAMI-VI was utilized for a reanalysis [63] of the MAMI-I and MAMI-IV data. Compared to the results of the original analysis, the extracted values for the $\alpha_E$ were slightly reduced, but the change was not considerable so as to eliminate the observed structure in the electric GP.

The most recent measurements involve the VCS program in Hall C at the Thomas Jefferson National Accelerator Facility. The first stage of this effort (the VCS-I experiment) [63] acquired data in the region $Q^2$ = 0.28 $GeV^2$ to 0.40 $GeV^2$, aiming for high-precision measurements in the region where the puzzling structure of $\alpha_E$ ($Q^2$) [56, 60] was previously reported. The experiment capitalized on the unique capabilities of the experimental setup in Hall C at Jefferson Lab that allowed us to conduct measurements of the scalar GPs with unprecedented precision. Cross-section measurements were conducted for azimuthally symmetric kinematics in the photon angle, i.e., for ($\phi_{\gamma^*\gamma}, \pi - \phi_{\gamma^*\gamma}$), since the measurement of the azimuthal asymmetry in the cross-section enhances the sensitivity in the extraction of the polarizabilities and suppresses part of the systematic uncertainties. Moreover, the ep→ep$\pi^0$ reaction was measured simultaneously with the ep→ep$\gamma$ reaction. The cross-section of the pion–electroproduction process is well understood in this kinematic regime, and its measurement offers a stringent, real-time normalization control for the measurement of the ep→ep$\gamma$ cross-section. Overall, a significant improvement was made in the precision of the extracted generalized polarizabilities compared to previous measurements.

The VCS-I data were acquired in Hall C of Jefferson Lab. Electrons with energies of 4.56 GeV at a beam current of up to 20 $\mu A$ were produced by the Jefferson Lab Continuous Electron Beam Accelerator Facility (CEBAF) and were scattered from a 10-cm-long liquid-hydrogen target. The Super High Momentum Spectrometer (SHMS) and the High Momentum Spectrometer (HMS) of Hall C were used to detect, in coincidence, the scattered electrons and recoil protons, respectively (see Figure 6). Both spectrometers are equipped with similar detector packages, including a set of scintillator planes that were used to form the trigger and provide time-of-flight information and a pair of drift chambers used for tracking. The coincidence time was determined as the difference in the time-of-flight between the two spectrometers, accounting for path-length variation corrections from the central trajectory and for the individual start times. The experimental setup offered a resolution of ~1 ns (FWHM) in the coincidence timing spectrum. Random coincidences were subtracted using the side (accidental) bands of the coincidence time spectrum. The events of the exclusive reaction ep → ep$\gamma$ were identified from the missing-mass reconstruction through a selection cut around the photon peak in the missing-mass-square spectrum. Data were taken with an empty target in order to account for the background contributions from the target walls. Elastic scattering measurements with a proton target were performed throughout the experiment for calibration and normalization studies.

The measurement of the absolute VCS cross-section, $\sigma \equiv d^5\sigma/dE'_e d\Omega'_e d\Omega_{cm}$, requires the determination of the five-fold solid angle, where $dE'_e$ and $d\Omega'_e$ are the differential energy and solid angle of the scattered electron in the laboratory frame, respectively, and $d\Omega_{cm}$ is the differential solid angle of the photon in the center-of-mass frame. The experimental acceptance was calculated with the Hall C Monte Carlo simulation program, SIMC, which integrates the beam configuration, target geometry, spectrometer acceptances, resolution effects, energy losses, and radiative corrections. The cross-section of the $ep \to ep\gamma$ process receives contributions from the photon that is emitted by either the lepton, i.e., the BH process, or by the proton, the fully virtual Compton scattering (FVCS) process. The FVCS amplitude can, in turn, be decomposed into a Born contribution, with the intermediate state being the nucleon, and a non-Born contribution that carries the physics of interest and is parametrized by the GPs. The BH and the Born-VCS contributions are well known and calculable in terms of the proton electromagnetic form factors. The GPs were extracted from the measured cross-sections through a fit that employs the DR model [21, 29, 30] for VCS. In DR formalism, the two scalar GPs enter unconstrained and can be adjusted as free parameters, while the proton electromagnetic form factors are introduced as input. The experimental cross-sections were compared to the DR model predictions for all possible values for the two GPs, and $\alpha_E$ ($Q^2$) and $\beta_M(Q^2)$ were fitted by $\chi^2$ minimization. The extracted electric and magnetic GPs from the VCS-I experiment are shown in Figure 7. The measurements suggest a local enhancement of $\alpha_E$ ($Q^2$) in the measured region, at the same $Q^2$ as previously reported by [54, 55, 60], but with a smaller magnitude than what was originally suggested. The $Q^2$-dependence of the electric GP was explored using two types of approaches [64], namely, methods that use traditional fits to the data using predefined functional forms and methods that are based on data-driven techniques that assume no direct underlying functional form. Both methods point to a $Q^2$-dependence for $\alpha_E(Q^2)$ that is consistent with the presence of a structure in the measured region, in sharp contrast with the current theoretical understanding that suggests a monotonic dependence of $\alpha_E(Q^2)$ with $Q^2$. For $\beta_M(Q^2)$, the results point to a smooth $Q^2$-dependence and the near-cancellation of the paramagnetic and the diamagnetic contributions in the proton at ~ $Q^2$ = 0.4 GeV$^2$. A comparison with the theory predictions of BChPT [46], NRQCM [32], LSM [37], ELM [36] and DR [21, 29, 30] is shown in Figure 7. As shown in the figure, the theoretical predictions for the GPs vary notably, and the experimental results impose strict constraints to the theory.

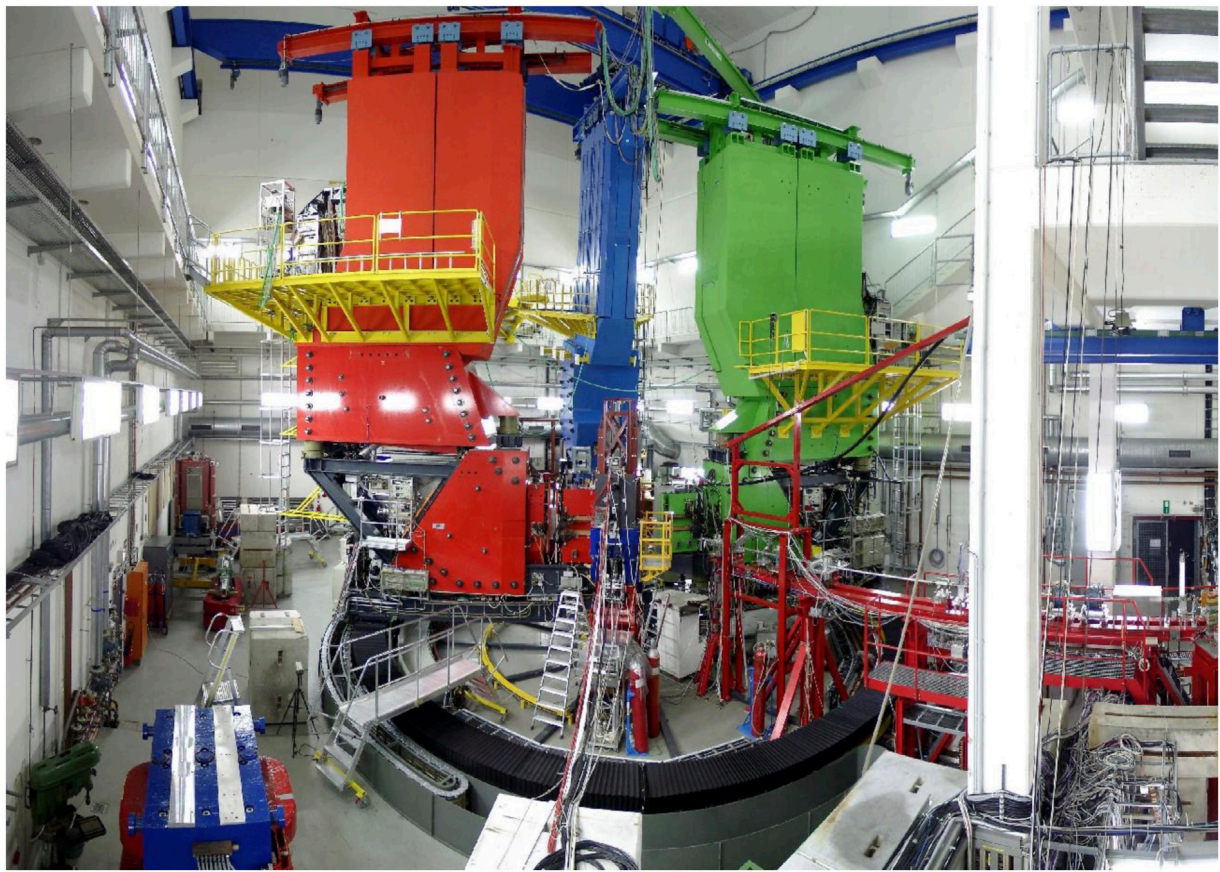

FIGURE 4
Three magnetic spectrometer setup of A1 at MAMI. The photo was provided by M. Weis, PhD thesis, University of Mainz, 2003.

## 3.2 Spin asymmetry measurements

The VCS experimental program has been extended beyond the unpolarized cross-section measurements to measurements of beam single-spin asymmetries (SSAs) and double-spin asymmetries (DSAs). Here, the goal is to study observables beyond the two scalar GPs, including the four lowest-order spin-dependent GPs. This effort has so far involved a couple of exploratory beam SSA and DSA experiments that turned out to be challenging and offered limited sensitivity to the observables of interest.

The beam single-spin asymmetry in VCS [65] was introduced to access non-trivial phases of QCD and test the diquark model predictions. The observable of interest is the asymmetry $(d\sigma^+ - d\sigma^-)/(d\sigma^+ + d\sigma^-)$, where $d\sigma^+$ and $d\sigma^-$ stand for the photon electroproduction cross-sections with a longitudinally polarized electron beam of helicity $+\frac{1}{2}$ and $-\frac{1}{2}$, respectively. The numerator of the asymmetry is equal to $\mathcal{I}m(T^{VCS}) \cdot \mathcal{R}e(T^{VCS} + T^{BH})$ and illustrates that the beam SSA is proportional to the imaginary part of the VCS amplitude. Consequently, one must measure above the pion-production threshold to access this asymmetry. Furthermore, the asymmetry vanishes at azimuthal angles of 0° and 180°, so one must measure at out-of-plane kinematics. One can distinguish the two terms that drive the beam SSA numerator. The first term, $\mathcal{I}m(T^{VCS}) \cdot \mathcal{R}e(T^{VCS})$, is exclusively due to the VCS contribution and measures the relative phase between longitudinal and transverse virtual Compton helicity amplitudes. The second one, $\mathcal{I}m(T^{VCS}) \cdot \mathcal{R}e(T^{BH})$, is an interference term that measures the relative phases between the VCS and BH amplitudes. In kinematics where the BH dominates the cross-section, this interference amplifies the VCS contribution and enhances the asymmetry. The MAMI-II experiment [66, 67] measured the beam SSA in the first resonance region at $Q^2$ = 0.35 GeV$^2$, covering out-of-plane c.m. angles up to 40°. The SSA was small, less than 10%, and the measurement was of a limited precision. The data illustrated a good agreement with the DR calculation that adopts the $(\gamma^{(*)}N \to \pi N)$ multipoles of the MAID [68, 69] analysis, but in the kinematics of the experiment, the DR calculation had little sensitivity to the GPs. On the other hand, the measurement had good sensitivity to the two longitudinal multipoles $S_{1+}$ and $S_{0+}$ in the $(p\pi^0)$ channel that are involved in the $(p\pi^0)$ intermediate state of the VCS process.

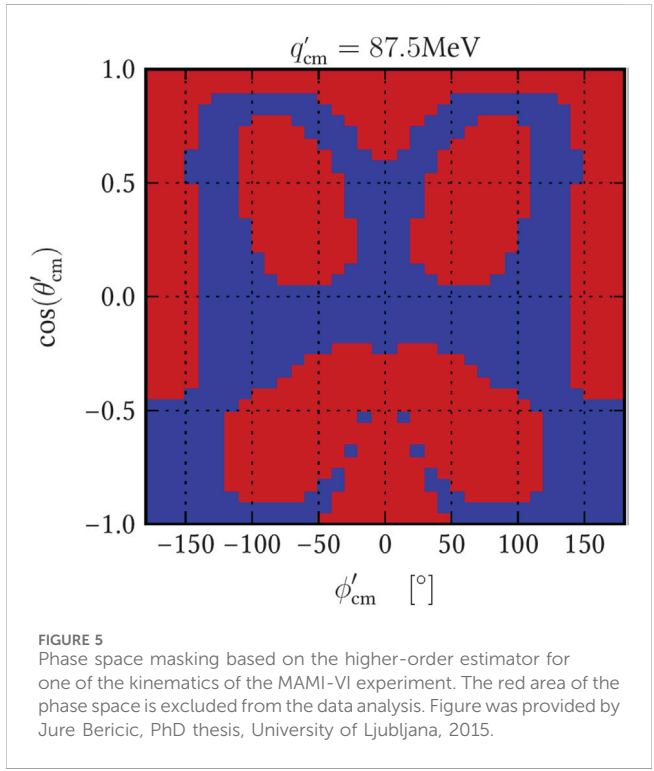

FIGURE 5
Phase space masking based on the higher-order estimator for one of the kinematics of the MAMI-VI experiment. The red area of the phase space is excluded from the data analysis. Figure was provided by Jure Bericic, PhD thesis, University of Ljubljana, 2015.

The theoretical context of the doubly polarized VCS has been studied by [20, 35]. The observables, i.e., the doubly polarized cross-sections or asymmetries, require polarization on both the leptonic and hadronic parts. Experimentally, the double-polarization observable is determined via Eq. 17

$$\mathcal{P}_i^{cm} = \frac{d^5\sigma(h,\hat{i}) - d^5\sigma(h,-\hat{i})}{d^5\sigma(h,\hat{i}) + d^5\sigma(h,-\hat{i})}, \quad (17)$$

where $\hat{i} = x, y, z$ is the c.m. axis for the recoil proton polarization component, $h = \pm\frac{1}{2}$ is the beam helicity, and $d^5\sigma(h,\hat{i})$ is the doubly polarized $(ep \to e'p'\gamma)$ cross-section. The LET expansion for the double-polarization observables [20, 35], which is valid below the pion threshold, leads to Eq. 18

$$\mathcal{P}_i^{cm} = \frac{\Delta d^5\sigma^{BH+B} + \phi q_{cm}'\Delta\mathcal{M}^{nB}(h,\hat{i}) + \mathcal{O}(q_{cm}'^2)}{2d^5\sigma}, \quad (18)$$

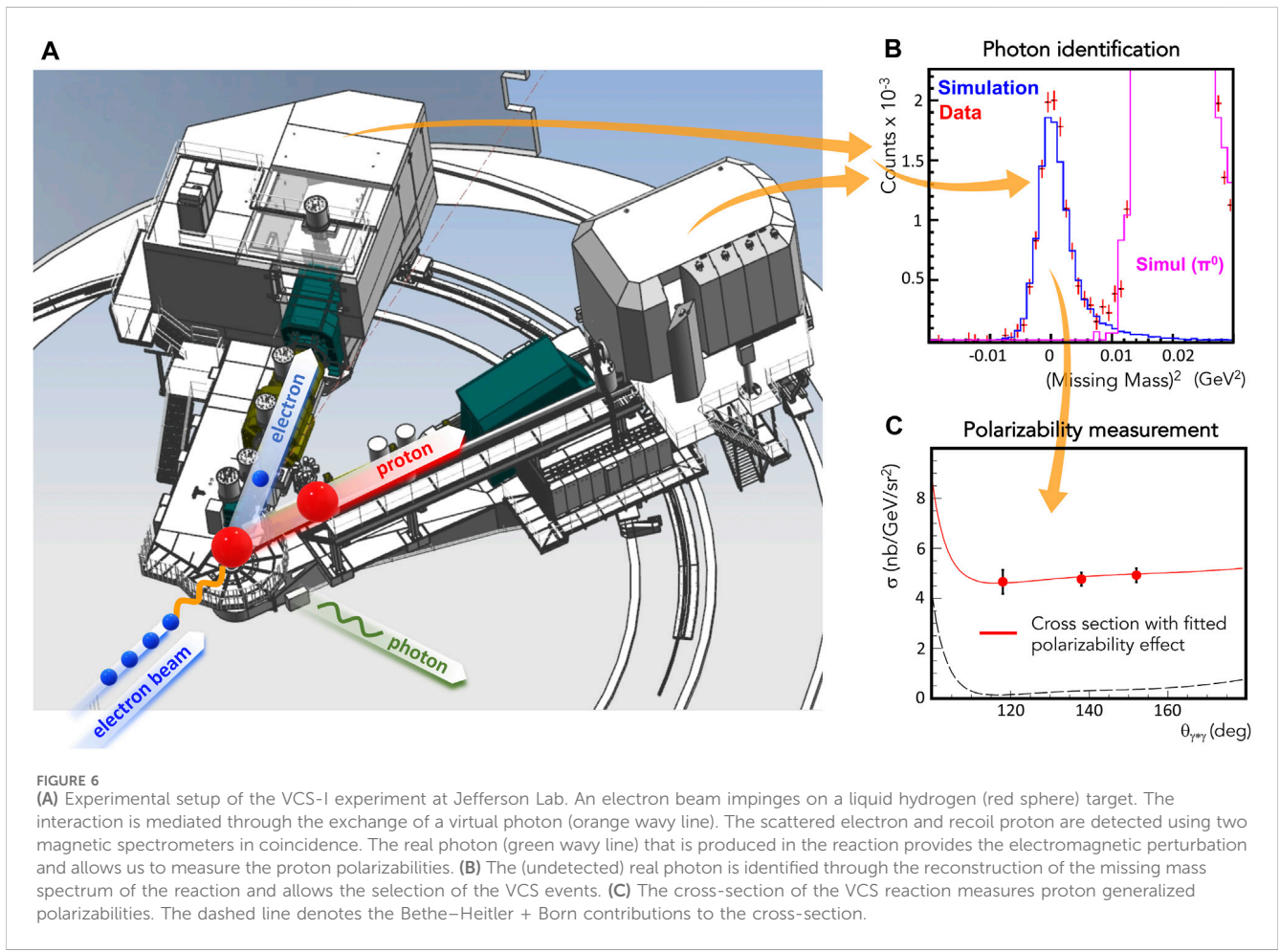

FIGURE 6
(A) Experimental setup of the VCS-I experiment at Jefferson Lab. An electron beam impinges on a liquid hydrogen (red sphere) target. The interaction is mediated through the exchange of a virtual photon (orange wavy line). The scattered electron and recoil proton are detected using two magnetic spectrometers in coincidence. The real photon (green wavy line) that is produced in the reaction provides the electromagnetic perturbation and allows us to measure the proton polarizabilities. (B) The (undetected) real photon is identified through the reconstruction of the missing mass spectrum of the reaction and allows the selection of the VCS events. (C) The cross-section of the VCS reaction measures proton generalized polarizabilities. The dashed line denotes the Bethe−Heitler + Born contributions to the cross-section.

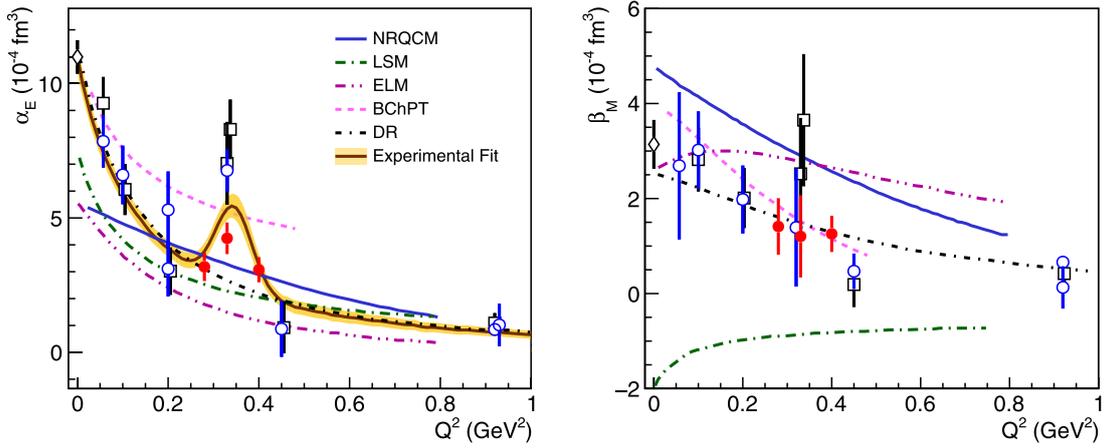

FIGURE 7
Electric and magnetic generalized polarizability measurements. The VCS-I results [64] are shown as red-circles, and the world data [14, 54–56, 58, 60–62] are presented as open symbols. The results from the dispersion-relation fits and the low-energy expansion fits are shown as circles and boxes, respectively. The theoretical predictions of BChPT [46], NRQCM [32], LSM [37], ELM [36], and DR [21, 29, 30] are also shown.

where $\Delta d^5\sigma^{BH+B}$ is the difference between the doubly polarized cross-sections $d^5\sigma^{BH+B}(h,\hat{i}) - d^5\sigma^{BH+B}(h,-\hat{i})$, $d^5\sigma$ is the unpolarized ($ep \to e'p'\gamma$) cross-section, and $(\phi q'_{cm})$ is a phase-space factor. The non-Born terms $\Delta \mathcal{M}^{nB}$ are linear combinations of the VCS structural functions $P_{LT}^\perp$, $P_{TT}^\perp$, $P_{LT}'^\perp$, $P_{LT}'^\perp$, $P_{LT}^z$, and $P_{LT}'^z$, which can be expressed as linear combinations of the six GPs, namely, the two scalar and the four spin-dependent GPs. More detailed formulas are given in the studies by [20, 35]. Double-polarization observables were explored in the MAMI-IV experiment, at kinematics similar to those of the MAMI-I experiment. The beam was longitudinally polarized, and a focal-plane polarimeter (FPP) was used to measure the recoil proton transverse polarization components. From the double-polarization observables, the structural function $P_{LT}^\perp$ was extracted for the first time. $P_{LT}^\perp$ is a linear combination of the structural functions $P_{LL}$ and $P_{TT}$, where $P_{LL}$ is proportional to the electric GP and $P_{TT}$ is a combination of the two spin GPs $P^{(M1,M1)1}$ and $P^{(L1,M2)1}$. The extracted value $P_{LT}^\perp = (-15.4 \pm 3.3_{stat}\ ^{+1.5}_{-2.4\ syst})$ GeV$^{-2}$ [70] is larger in magnitude than most model predictions. This exploratory double-polarization experiment turned out to be more challenging than the unpolarized experiments, and the measured double-spin asymmetry was less sensitive than expected to the GPs. In order to disentangle all six lowest-order GPs, one will require an extensive experimental program, with more data that span a wider range of the kinematical variables.

# 4 Induced polarization in the proton

The measurements of the generalized polarizabilities provide access to the spatial deformation of the quark distributions in the proton subject to the influence of an external electromagnetic field [71]. Here, the method follows an extension of the formalism to extract the light-front quark charge densities [72] from the proton form factor data. The induced polarization in the proton from the GP world data was extracted by [64]. In this work, an accurate parametrization of the polarizabilities was first derived from a fit to the experimental data, and from that, the induced polarization in the proton, $P_0$, defined by Eqs 19, 20 was extracted based on the formalism discussed by [71], i.e.,

$$\vec{P}_0(\vec{b}) = \hat{b} \int_0^\infty \frac{dQ}{(2\pi)} Q J_1(bQ) A(Q^2), \quad (19)$$

where $\vec{b}$ is the transverse position, $b = |\vec{b}|$, $\hat{b} = \vec{b}/b$, and $J_1$ is the first-order Bessel function. $A$ is a function of the GPs:

$$A = -(2M_N) \sqrt{\tau} \sqrt{\frac{3}{2}} \sqrt{\frac{1+2\tau}{1+\tau}}$$
$$\times \left\{ -P^{(L1,L1)0} + \frac{1}{2} P^{(M1,M1)0} - \sqrt{\frac{3}{2}} P^{(L1,L1)1} \right.$$
$$\left. -\sqrt{\frac{3}{2}} (1+\tau) \left[ P^{(M1,M1)1} + \sqrt{2}\ (2M_N \tau) P^{(L1,M2)1} \right] \right\}.$$
(20)

The GPs are expressed in the typical multipole notation $P^{(\rho'l',\rho l)S}$ [26], where $\rho$ ($\rho'$) refers to the Coulomb/electric ($L$) or magnetic ($M$) nature of the initial (final) photon, $l$ ($l' = 1$) is the angular momentum of the initial (final) photon, and $S$ differentiates between the spin-flip ($S = 1$) and non-spin-flip ($S = 0$) transition at the nucleon side. $\tau \equiv Q^2/(4M_N^2)$, with $M_N$ being the nucleon mass. In calculating the $A$ function, the two scalar GPs are defined using Eqs 5, 6, and the spin GPs are fixed by the dispersion relations. For the asymptotic part of $\alpha_E(Q^2)$, one uses the following parametrization of Eq. 21 that is derived from an experimental fit to the world data:

$$\alpha_E(Q^2) = \left[ p0 * e^{-0.5*\left(\frac{Q^2-p1}{p2}\right)^2} + \frac{1}{(p3 + Q^2/p4)^2} \right] (fm^3), \quad (21)$$

with $p0 = (30.4 \pm 6.1)10^{-5}\ (fm^3)$, $p1 = 0.345 \pm 0.008\ (GeV^2)$, $p2 = 0.040 \pm 0.003\ (GeV^2)$, $p3 = 34.217 \pm 1.136\ (fm^{-\frac{3}{2}})$, and $p4 = 0.014 \pm 0.002\ (GeV^2 fm^{\frac{3}{2}})$. For $\beta_M(Q^2)$, the world data are described accurately using the DR model [21, 29, 30], following a single-dipole behavior for the unconstrained part of the scalar GPs with a mass scale parameter of $\Lambda_\beta = 0.5\ GeV$. The derived induced polarization as a function of the transverse position of the proton is shown in Figure 8. The distribution follows a change of sign of ~0.25 fm and exhibits a secondary maximum in the amplitude of ~0.35 fm.

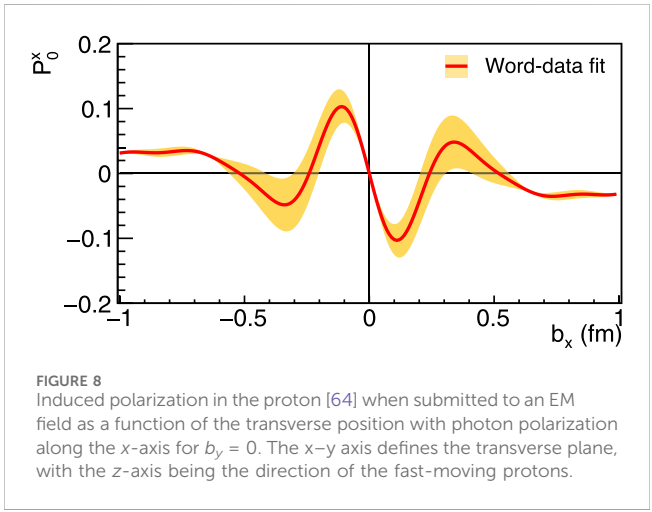

FIGURE 8
Induced polarization in the proton [64] when submitted to an EM field as a function of the transverse position with photon polarization along the x-axis for $b_y = 0$. The x–y axis defines the transverse plane, with the z-axis being the direction of the fast-moving protons.

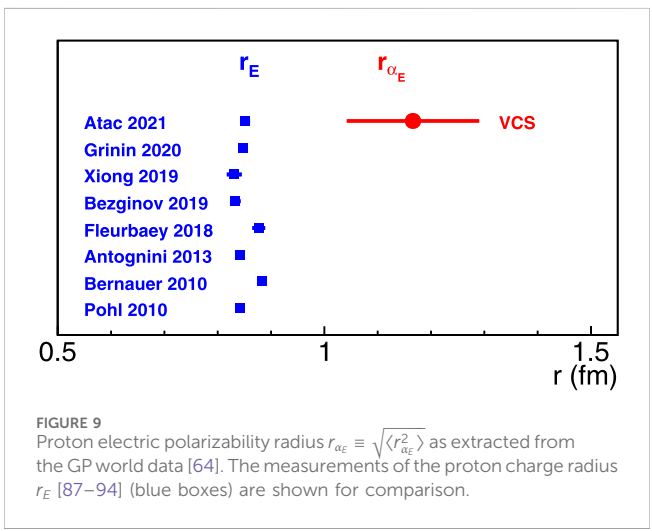

FIGURE 9
Proton electric polarizability radius $r_{\alpha_E} \equiv \sqrt{\langle r^2_{\alpha_E} \rangle}$ as extracted from the GP world data [64]. The measurements of the proton charge radius $r_E$ [87–94] (blue boxes) are shown for comparison.

# 5 Electric and magnetic polarizability radii

A primary measure that quantifies the extension of a spatial distribution is the mean square radius. The mean square electric polarizability radius of the proton $\langle r^2_{\alpha_E} \rangle$ can be determined from measurements of the VCS reaction since it is related to the slope of $\alpha_E(Q^2)$ at $Q^2 = 0$ as defined by Eq. 22

$$\langle r^2_{\alpha_E} \rangle = \frac{-6}{\alpha_E(0)} \cdot \frac{d}{dQ^2}\alpha_E(Q^2)|_{Q^2=0}. \quad (22)$$

The first $\langle r^2_{\alpha_E} \rangle$ extraction became possible with the MIT Bates GP measurements at $Q^2 = 0.057$ $GeV^2$ [58]. In this work, the mean square electric polarizability radius was determined from a dipole fit to the RCS and the MIT Bates data points, giving $\langle r^2_{\alpha_E} \rangle = 1.95 \pm 0.33$ $fm^2$ [58]. A more reliable extraction was done recently, taking into account a more extensive dataset from modern experiments in tandem with a comprehensive consideration of the fitted functions and the fitting range. Li et al. (2022a) conducted the $\langle r^2_{\alpha_E} \rangle$ extraction using a variety of functional forms that can fit the data, i.e., combinations of polynomial, dipole, Gaussian, and exponential functions. The fits were explored in two groups: one group considered the full $Q^2$ range and the second focused on a limited range at low-$Q^2$ that does not include the $\alpha_E$ anomaly, i.e., within $Q^2 = [0,0.28]$ $GeV^2$. For the experiments where the polarizabilities were derived by both the dispersion relations and the low-energy expansion analysis, the variance of the two results is treated as a model uncertainty for each data point. For each group of fits, the final value for $\langle r^2_{\alpha_E} \rangle$ is determined from the weighted average of the results of the individual fits. The uncertainty of $\langle r^2_{\alpha_E} \rangle$ receives contributions from two terms, the uncertainty of the weighted average and the weighted variance of the individual fit results, which effectively reflect the model dependence on the choice of the fitted parametrization. The final result is then derived from the average of the two group values, with their spread accounted for as a model uncertainty. The extraction of the polarizability radius is sensitive to the value of the static ($Q^2 = 0$) electric polarizability. If one considers the recent measurement from MAMI [14], $\alpha_E(0) = (10.99 \pm 0.16 \pm 0.47 \pm 0.17 \pm 0.34) 10^{-4}$ $fm^3$, the extracted value for the polarizability radius following the above analysis procedure is $\langle r^2_{\alpha_E} \rangle = 1.36 \pm 0.29$ $fm^2$. We note that the MAMI measurement [14] is in excellent agreement with the $\alpha_E(0)$ PDG value. Nevertheless, a recent experiment at HIGS [15] points to a tension compared to the MAMI measurement and reports a higher value of $\alpha_E(0) = (13.8 \pm 1.2 \pm 0.1 \pm 0.3) 10^{-4}$ $fm^3$, albeit with a higher experimental uncertainty. If one adopts the HIGS value for $\alpha_E(0)$, one derives the polarizability radius $\langle r^2_{\alpha_E} \rangle = 1.67 \pm 0.50$ $fm^2$. In both cases, extracted $\langle r^2_{\alpha_E} \rangle$ is considerably larger than the mean square charge radius of the proton, i.e., $\langle r^2_E \rangle \sim 0.7$ $fm^2$ [6,73] as shown in Figure 9. The dominant contribution to this effect is expected to arise from the deformation of the mesonic cloud in the proton under the influence of an external EM field.

For the extraction of the mean square magnetic polarizability radius from the magnetic polarizability measurements, one follows a procedure that is equivalent to the extraction of $\langle r^2_{\alpha_E} \rangle$ since the magnetic polarizability radius expression (Eq. 23) is similar, i.e.,

$$\langle r^2_{\beta_M} \rangle = \frac{-6}{\beta_M(0)} \cdot \frac{d}{dQ^2}\beta_M(Q^2)|_{Q^2=0}. \quad (23)$$

The reported values for $\beta_M(0)$ from the MAMI [14] and HIGS [15] experiments exhibit tensions similar to those were reported for $\alpha_E(0)$. If one adopts the MAMI measurement, $\beta_M(0) = (3.14 \pm 0.21 \pm 0.24 \pm 0.20 \pm 0.35) 10^{-4}$ $fm^3$, that is of higher precision and in good agreement with the PDG value, one derives $\langle r^2_{\beta_M} \rangle = 0.63 \pm 0.31$ $fm^2$ [64]. If instead one adopts the HIGS measurement, $\beta_M(0) = (0.2 \pm 1.2 \pm 0.1 \pm 0.3) 10^{-4}$ $fm^3$, then the magnetic polarizability radius is consistent with 0, i.e., $\langle r^2_{\beta_M} \rangle = 0.20 \pm 0.36$ $fm^2$.

# 6 Future experiments and prospects

## 6.1 Unpolarized cross-section measurements

The next phase of the VCS program at Jefferson Lab will involve the VCS-II experiment [74]. Cross-section measurements of the VCS reaction with an unpolarized electron beam will use the VCS-I experimental setup in Hall

Sparveris10.3389/fphy.2024.1426128

C to extend the kinematic coverage and improve the precision of the measured GPs. Data will be acquired for 62 days of beam-on-target, with an $E_\circ$ = 1.1 $GeV$ and 2.2 $GeV$ electron beam at I = 75 $\mu$A impinged on a 10-cm liquid-hydrogen target. Measurements will be taken within the region $Q^2$ = 0.05 $(GeV/c)^2$ to $Q^2$ = 0.50 $(GeV/c)^2$ to provide high-precision data combined with a fine mapping as a function of $Q^2$. The experiment will explore $\alpha_E$ with a set of measurements that are all uniform regarding their systematic uncertainties. This will make it possible to identify the shape of the observed structure in terms of electric polarizability more reliably and with high precision. Moreover, a systematic set of measurements will be conducted in kinematics, where the suggested structure in $\alpha_E$ emerges in an anti-diametric way in the VCS cross section. As shown in Figure 10, the sensitivity of the VCS cross-section to $\alpha_E$ undergoes a crossing point and reverses for the two wings of the resonance. Targeted measurements on both wings of the resonance will allow to decouple the observation of the non-trivial structure in the polarizability from the influence of experimental systematic uncertainties. The precision of the $\beta_M$ measurements will be further improved to allow a good handle on the interplay between the diamagnetic and paramagnetic contributions in the nucleon. The projected VCS-II measurements are shown in Figure 11.

## 6.2 Beam-charge and beam-spin asymmetry measurements

The recent measurements for the GPs highlight the need to access these quantities by employing alternative experimental methods. This will allow us to provide an independent confirmation, in particular for the observed structure in the electric polarizability. So far, the experimental measurements of the proton generalized polarizabilities have used unpolarized electron beams. The only exception is that of the exploratory MAMI measurement of beam-spin asymmetries that, as discussed in the previous section, had a weak sensitivity to polarizabilities. An alternative and powerful avenue to access the proton GPs beyond the unpolarized VCS measurements with an electron beam is presented by the use of polarized and positron beams [75]. The lepton beam charge ($e$) and polarization ($\lambda$) dependence of the $lp \rightarrow lp\gamma$ differential cross-section are given by Eq. 24

$$d\sigma^e_\lambda = d\sigma_{BH} + d\sigma_{VCS} + \lambda \, d\tilde{\sigma}_{VCS} + e \left( d\sigma_{INT} + \lambda \, d\tilde{\sigma}_{INT} \right), \quad (24)$$

where $d\sigma$ ($d\tilde{\sigma}$) represents the polarization-independent (dependent) contributions, which are even (odd) functions of the azimuthal angle $\phi$. $d\sigma_{INT}$ involves the real part of the VCS amplitude that contains the GP effects, while $d\tilde{\sigma}_{INT}$ is proportional to the imaginary part of the VCS amplitude, which does not depend on the GPs. Combining lepton beams of opposite charge and different polarizations enables the complete separation of the four unknown INT and VCS contributions. By employing unpolarized electron and positron beams, one can construct the unpolarized beam-charge asymmetry (BCA) $A^C_{UU}$ as defined in Eq. 25

$$\begin{aligned} A^C_{UU} &= \frac{(d\sigma^+_+ + d\sigma^+_-) - (d\sigma^-_+ + d\sigma^-_-)}{d\sigma^+_+ + d\sigma^+_- + d\sigma^-_+ + d\sigma^-_-} \\ &= \frac{d\sigma_{INT}}{d\sigma_{BH} + d\sigma_{VCS}}. \end{aligned} \quad (25)$$

With polarized lepton beams, on the other hand, one can construct the lepton beam-spin asymmetry (BSA), as defined by Eq. 26.

$$\begin{aligned} A^e_{LU} &= \frac{d\sigma^e_+ - d\sigma^e_-}{d\sigma^e_+ + d\sigma^e_-} \\ &= \frac{d\tilde{\sigma}_{VCS} + e d\tilde{\sigma}_{INT}}{d\sigma_{BH} + d\sigma_{VCS} + e \, d\sigma_{INT}}. \end{aligned} \quad (26)$$

The theoretical groundwork and the first study for the potential of this type of experiment were presented by [75]. It was illustrated that targeted measurements of un-polarized BCAs and polarized BSAs exhibit remarkable sensitivity to both scalar GPs. A combination of both types of asymmetries, such as in Eqs 27, 28

$$\begin{aligned} \tilde{A}_{VCS} &\equiv A^+_{LU}\left(1 + A^C_{UU}\right) + A^-_{LU}\left(1 - A^C_{UU}\right) \\ &= \frac{2d\tilde{\sigma}_{VCS}}{d\sigma_{BH} + d\sigma_{VCS}}, \end{aligned} \quad (27)$$

and

$$\begin{aligned} \tilde{A}_{INT} &\equiv A^+_{LU}\left(1 + A^C_{UU}\right) - A^-_{LU}\left(1 - A^C_{UU}\right) \\ &= \frac{2d\tilde{\sigma}_{INT}}{d\sigma_{BH} + d\sigma_{VCS}}, \end{aligned} \quad (28)$$

is powerful toward separating the contribution from the $d\tilde{\sigma}_{VCS}$ and $d\tilde{\sigma}_{INT}$ terms, offering not only sensitivity to the GPs but also a cross-check of the unitarity input in the dispersive formalism, as discussed by [75].

Plans for a first experiment with beam-charge and beam-spin asymmetry measurements at Jefferson Lab have been presented at JLab, PAC51 Letter-of-Intent LOI-12-23-001 [76]. For the unpolarized beam-charge asymmetries, the proposed experiment targets $Q^2$ = 0.35 $GeV^2$, the kinematics of high interest for the electric GP. The experiment will require a combination of electron and positron beam measurements at an energy of $E_\circ$ = 2.2 GeV. The SHMS and HMS magnetic spectrometers in Hall C will measure $e^{(+,-)}$ and p, respectively, while the range of center-of-mass energies that exhibit optimal sensitivity to the polarizabilities spans $W$ = 1,150 $MeV$–$W$ = 1,190 $MeV$. The measurements will access out-of-plane kinematics up to $\phi$ = 30$^o$, and $\theta_{\gamma^*\gamma}$ spans a range, as shown in Figure 12, for one bin in the center-of-mass energy. The experiment aims for a 1% statistical uncertainty for each of the (+,-) measurements, and the results will be limited by the systematic uncertainties. The experiment will require 1 week of beam-on-target with an electron beam at a current of $I \sim 50~\mu A$. The measurements with a positron beam will require more beam time since the accelerator cannot offer such a high level of beam current for the positrons. The beam time will depend on the final performance of the delivered positron beam, which is currently under development at Jefferson Lab. For an unpolarized positron beam, it is expected at the level of a few $\sim \mu A$. In an optimistic scenario where one can achieve 5 $\mu A$, approximately 10 weeks of beam-on-target will be needed for the experiment. The projected measurement for $\alpha_E$ is shown in Figure 12.

The use of a polarized beam offers one more alternative path for GP measurements. Here, the goal is to measure BSAs, which can be done independently with either an electron or a positron beam. Such measurements with a polarized electron beam can become readily available at Jefferson Lab. A measurement is considered again at $Q^2 = 0.35$ $GeV^2$ with an electron beam energy of $E_\circ = 2.2$ $GeV$. Here, the sensitivity to the polarizabilities is enhanced at higher center-of-mass energies, in the range $W = 1210$ $MeV$–$W = 1250$ $MeV$. The projected results for the proposed measurements, considering ~2 weeks of beam-on-target with an electron beam of $I = 70$ $\mu A$ at 85% polarization, are shown in Figure 13. The beam-spin asymmetry allows us to suppress many of the systematic uncertainties, and the statistics become the limiting factor for these measurements. One can thus run additional beam time in order to further reduce the experimental uncertainties that are shown in Figure 13. The scenario of a BSA measurement with a positron beam has also been considered. For a group of similar measurements (i.e., at the same kinematics and with the same precision as the electron BSA measurements), one can, in principle, achieve an equally competitive extraction of the polarizabilities compared to the measurements with an electron beam. Nevertheless, the limitation here involves the beam time that is needed for the measurements, since the beam current for a polarized positron beam at Jefferson Lab will be significantly suppressed. Considering a beam current of $I \sim 50$ $nA$ and a beam polarization of 60%, one will need a beam time that is three orders of magnitude higher compared to the measurements with an electron beam. As such, this path is not viable based on the expected performance of a future positron beam at JLab.

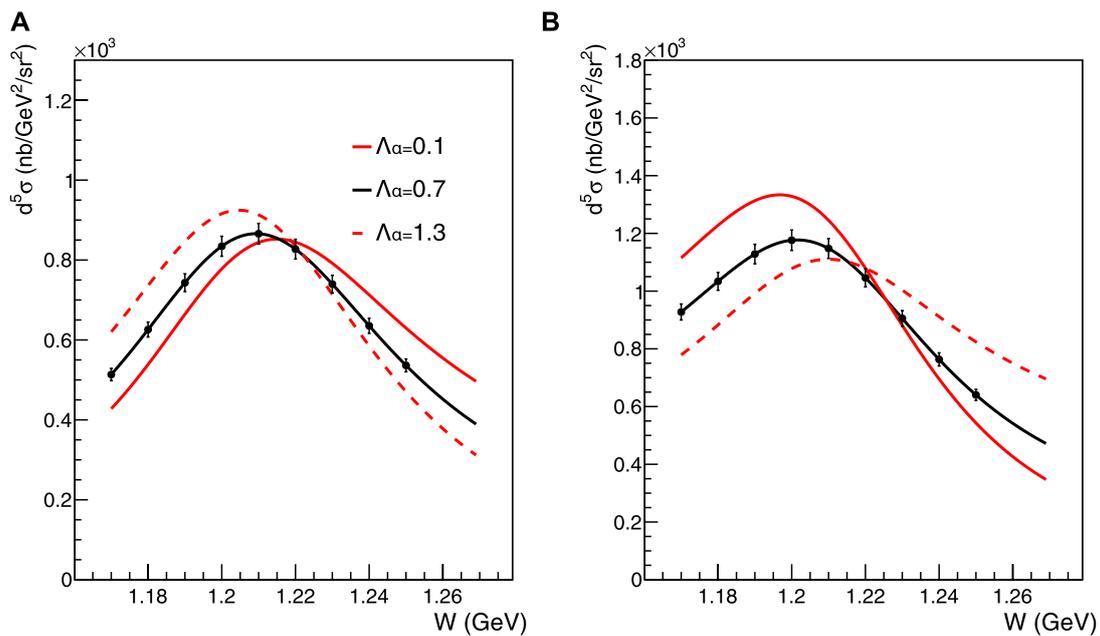

FIGURE 10
Projected cross-section measurements of the VCS-II experiment. The W-dependence of the cross-section is shown at $Q^2 = 0.35$ $(GeV/c)^2$ for $\theta_{\gamma^*\gamma} = 140°$, and $\phi_{\gamma^*\gamma} = 0°$ **(A)**, and $180°$ **(B)**. The different curves illustrate the sensitivity of the cross section to $\alpha_E$, within the range $\Lambda_\alpha = 0.1$–$1.3$.

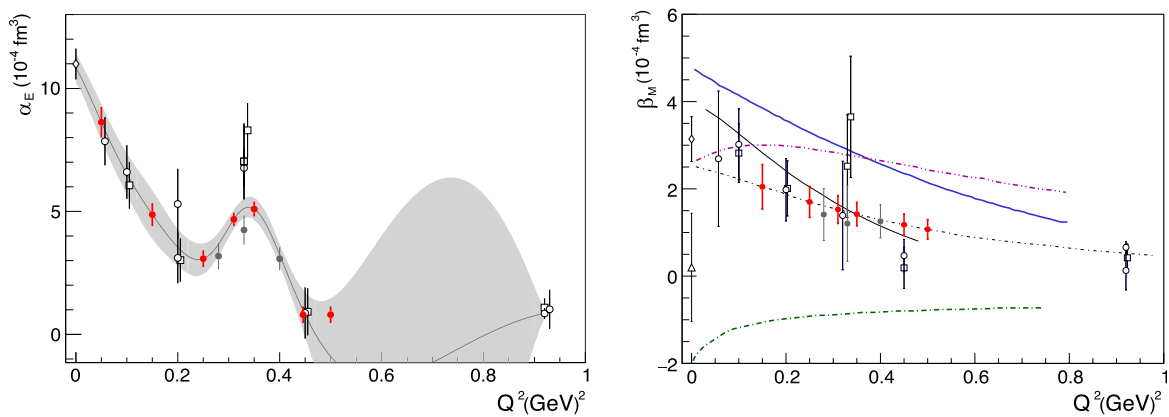

FIGURE 11
The VCS-II-projected measurements for $\alpha_E$ and $\beta_M$ are shown as red solid points. The world data are shown as open symbols, with the exception of the VCS-I results that are indicated by filled gray circles. The gray band in the electric GP plot indicates the data-driven extraction of the polarizability [64].

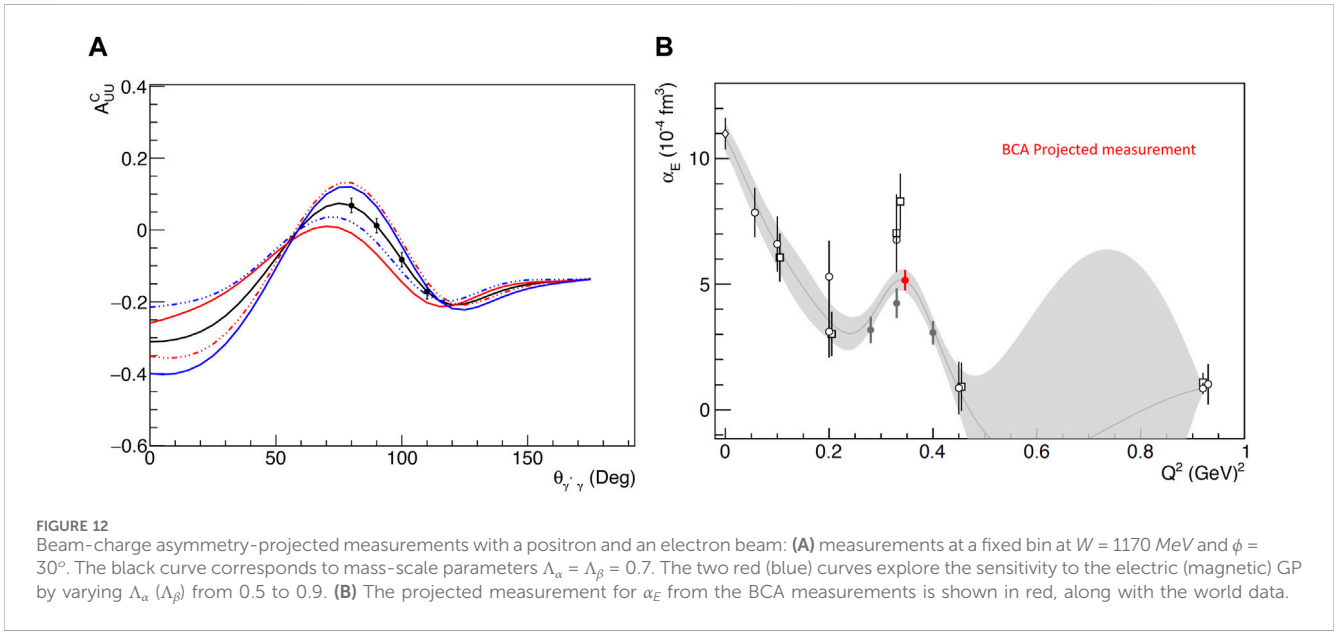

FIGURE 12
Beam-charge asymmetry-projected measurements with a positron and an electron beam: **(A)** measurements at a fixed bin at $W = 1170$ *MeV* and $\phi = 30°$. The black curve corresponds to mass-scale parameters $\Lambda_\alpha = \Lambda_\beta = 0.7$. The two red (blue) curves explore the sensitivity to the electric (magnetic) GP by varying $\Lambda_\alpha$ ($\Lambda_\beta$) from 0.5 to 0.9. **(B)** The projected measurement for $\alpha_E$ from the BCA measurements is shown in red, along with the world data.

## 6.3 Conclusion and outlook

The Compton scattering process provides a unique tool to study the electromagnetic polarizability mechanisms in the proton. The generalization of the polarizabilities to non-zero four-momentum transfer extends the applicability of these measurements to the study of the spatial distribution of the polarization densities in the proton, offering a unique insight into the underlying dynamics of the fundamental quark and gluon constituents. The experimental signal of interest is very small, and the measurements come with significant challenges. Three decades of virtual Compton scattering experiments have led to remarkable progress in the measurement of these fundamental properties of the proton. The experimental progress has been enabled by the advances in the theoretical front and by the close synergy of the theoretical and experimental efforts. The recent experiments illustrated that we accomplished a significant improvement in the precision of the measurements compared to the early VCS experiments. The electric and magnetic GPs of the proton are now understood with a good level of precision. The magnetic polarizability signal is more challenging to extract, considering the cancellation of the paramagnetic and diamagnetic contributions in the proton. Nevertheless, there is room for improvement, and upcoming experiments will add further to our understanding of both scalar GPs. A question remains regarding the observed structure in the electric GP that the theory is not able to explain. This structure is suggested by three independent experimental measurements [54, 55, 60, 64]. The world data suggest a deviation from a monotonic fall-off with $Q^2$ at intermediate momentum transfers within ~3σ [64]. This allows for a ~1% probability that the observation is coincidental.

Here, one may consider the possibility that some of the systematic uncertainties associated with the analysis of the VCS measurements may have been underestimated and that they are not under control within the reported level. Toward that end, it becomes important to come forward with different methods of measurement for the GPs. This was proposed recently [76] by pursuing beam-charge and beam-spin asymmetry measurements. If the observed structure in the electric GP is confirmed by a new experiment that involves a different framework for the extraction of the polarizabilities, then a definitive answer will have been provided on the existence of the effect. A measurement of beam-spin asymmetries with an electron beam can be done at Jefferson Lab with the existing capabilities of the laboratory. A beam-charge asymmetry measurement will become possible in the near future once a positron beam at Jefferson Lab becomes available.

The experimental measurements of the GPs serve as high-precision benchmark data for the theory. They offer guidance to the theoretical calculations and our understanding of the underlying dynamics in the nucleon. Considering the size of the theoretical uncertainties, the experiment is currently ahead of the theory in terms of precision. The chiral effective field theory offers a quantum field theory with the Lagrangian written in terms of hadronic fields, in contrast to QCD, which is written in terms of quark and gluon fields and allows a QCD description of low-energy phenomena. The most recent work involves the next-to-leading order BChPT calculation [46] that is shown in Figure 14. The theory is in good agreement with the experimental data, but as shown in the figure, the results come with relatively large uncertainties, pointing to the need for the next-order calculation. These calculations are expected to become available in the near future. Regardless of the size of the theoretical uncertainties, BChPT prediction does not account for any irregularity in $\alpha_E(Q^2)$ that breaks the monotonic fall-off with $Q^2$. The lattice QCD (LQCD) allows an alternative path to calculate the polarizabilities through a space–time discretization of the theory based on the fundamental quark and gluon degrees of freedom, starting from the original QCD Lagrangian. The progress on that front is steady but still slow. These calculations are still limited to static polarizabilities. The LQCD is in agreement with the experimental data but within relatively large uncertainties. The recent LQCD calculation [77] at the physical pion mass reports for the electric polarizability $\alpha_E = (11.8 \pm 2.3 \pm 3.7)\ 10^{-4}\ fm^3$. By the end of the decade, one can expect that the LQCD uncertainties for the static polarizabilities will improve to a level of precision that is comparable to the experiment, and that the first LQCD calculations for the GPs will become available.

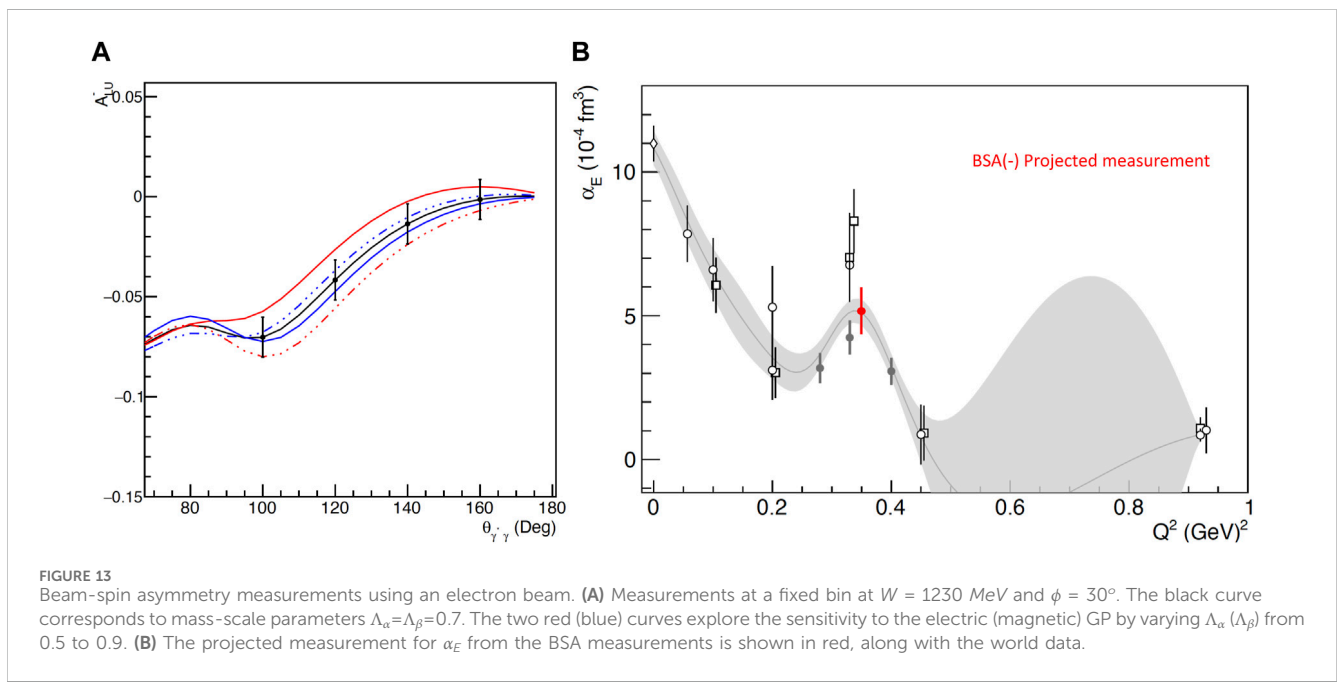

FIGURE 13
Beam-spin asymmetry measurements using an electron beam. **(A)** Measurements at a fixed bin at $W = 1230$ MeV and $\phi = 30°$. The black curve corresponds to mass-scale parameters $\Lambda_\alpha = \Lambda_\beta = 0.7$. The two red (blue) curves explore the sensitivity to the electric (magnetic) GP by varying $\Lambda_\alpha$ ($\Lambda_\beta$) from 0.5 to 0.9. **(B)** The projected measurement for $\alpha_E$ from the BSA measurements is shown in red, along with the world data.

While the chiral effective field theory and LQCD are making progress to provide refined calculations, parallel efforts focus on the study of potential mechanisms that could account for the observed structure in the $\alpha_E\ (Q^2)$. A recent study focused on the role of the resonance contributions to the generalized electric and magnetic nucleon polarizabilities, analyzed within the holographic QCD model [78]. Based on this calculation, as shown in Figure 15, the resonance contributions alone cannot account for the observed structure in $\alpha_E\ (Q^2)$. Nevertheless, there may be confusion because the type of GPs in the calculation involves only the symmetric GPs that are calculated in the forward VVCS and have to be distinguished from the ones in the VCS, as discussed by [46].

The article discussed the recent progress and developments of the VCS program and the proton EM generalized polarizabilities. Future experiments have the potential to extend the application of the VCS program beyond the study of the scalar GPs to study the spin-dependent GPs. Although the spin-dependent polarizability mechanisms do not allow for an intuitive and simplistic description of the underlying effects, their measurement can contribute constructively to the understanding of the nucleon structure. Here, we outline in brief some opportunities along this line of measurement. One prospect is presented through the measurement of the $P_{TT}$ structural function, which is a combination of two spin-dependent GPs, as indicated in Eq. 10. $P_{TT}$ appears in the LEX formalism (i.e., in Eqs 7, 8) in the $(P_{LL} - P_{TT}/\epsilon)$ term. Thus, one can extract $P_{TT}$ by conducting unpolarized VCS measurements at different values of $\epsilon$ for a common $Q^2$. The magnitude of $P_{TT}$ is expected to be small, according to DR and BChPT calculations, and this renders the extraction of the experimental signal challenging. Another path is to target a combination of unpolarized and polarized observables at a common value of $\epsilon$, as discussed by [79, 80], in order to measure different linear combinations of $P_{LL}$ and $P_{TT}$ and ultimately accomplish their separation. This method was first explored in the MAMI-IV experiment, but the correlation of the two structural functions from the two (polarized/unpolarized) measurements had a strong overlap and did not allow a good separation of the two terms. One can revisit this strategy, targeting an optimal combination of kinematics that will allow the extraction of $P_{LL}$ and $P_{TT}$. Additional opportunities can be presented in the future once the DR model is further developed to allow the spin-dependent GPs (currently predicted by the DR) to enter as free parameters in the calculation. This will empower the DR framework to explore the simultaneous extraction of the scalar and spin GPs by combining an extended set of VCS measurements and taking advantage of the enhanced sensitivity of the GPs in the Δ resonance region. Lastly, sum rules [81] have been derived in recent years, relating VCS, RCS, and doubly virtual Compton scattering (VVCS) at low-momentum transfers. [82], by generalizing the Gerasimov–Drell–Hearn sum rule [83–85] to finite photon virtuality, obtained two new model-independent relations, linking parameters that characterize different sectors of low-energy interactions between the nucleon spin structure and electromagnetic waves. These parameters are extracted from experimental information in VCS, RCS, and VVCS, i.e., they involve the generalized polarizabilities, the spin polarizabilities, the longitudinal–transverse polarizability, and the generalized GDH integral, while the nucleon form factors and the anomalous magnetic moments enter as additional input in the sum rules. These sum rules can be tested with experimental measurements and can be useful in constraining the low-energy spin structure of the nucleon.

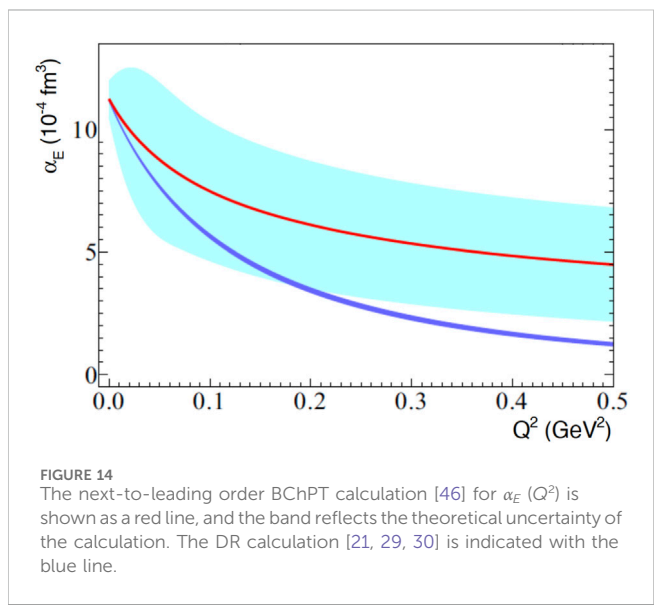

FIGURE 14
The next-to-leading order BChPT calculation [46] for $\alpha_E\ (Q^2)$ is shown as a red line, and the band reflects the theoretical uncertainty of the calculation. The DR calculation [21, 29, 30] is indicated with the blue line.

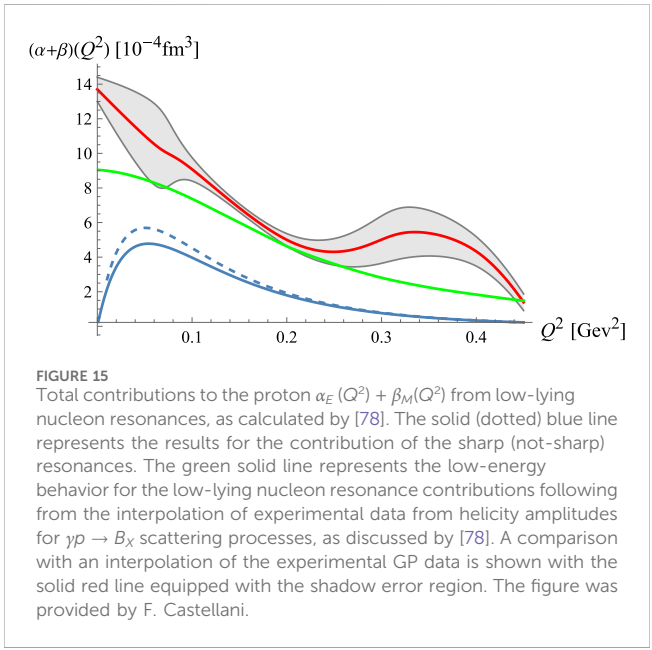

FIGURE 15
Total contributions to the proton $\alpha_E(Q^2) + \beta_M(Q^2)$ from low-lying nucleon resonances, as calculated by [78]. The solid (dotted) blue line represents the results for the contribution of the sharp (not-sharp) resonances. The green solid line represents the low-energy behavior for the low-lying nucleon resonance contributions following from the interpolation of experimental data from helicity amplitudes for $\gamma p \to B_X$ scattering processes, as discussed by [78]. A comparison with an interpolation of the experimental GP data is shown with the solid red line equipped with the shadow error region. The figure was provided by F. Castellani.

In summary, this article reviews our knowledge about a fundamental property of the proton, i.e., its response to an external electromagnetic field, which is described by the scalar polarizabilities and their generalization to non-zero momentum transfer. The recent progress of the last 5 years involves, for the most part, new experimental results. As such, the paper emphasizes the recent developments and findings of the virtual Compton scattering experiments and the discussion of future perspectives on the topic, which are discussed in tandem with the advances on the theory front. The first conclusion is that experimental progress is currently ahead of theory. The recent experiments have achieved a high level of precision in the measurement of the GPs and will soon be extended across the full range of the kinematical phase space. GPs are a key element of the proton structure, and they complement the information that is provided by the nucleon form factors, providing access to the spatial distribution of the polarization densities in the proton. The induced polarization as a function of the transverse position of the proton has been measured quite accurately, providing an insight into the spatial dependence of the polarizability effects. Fundamental characteristics of the system, such as the electric and magnetic polarizability radii, have been determined with good precision due to the recent experimental advances of the VCS experiments. A long-standing puzzle involves electric polarizability. A structure in $\alpha_E(Q^2)$ has been observed at intermediate momentum transfers, which breaks the monotonic dependence with $Q^2$. This observation cannot be explained by the theory at the moment. The effect has been confirmed by three independent experiments at the $3\sigma$ level. This leaves a possibility of ~1% that the observation may be coincidental, while one should consider the possibility that part of the systematic uncertainties of the VCS experiments may have been underestimated and should be revisited. A critical part of future experimental studies will involve the measurement and confirmation of this effect using a different method, as proposed by the beam-spin and beam-charge asymmetry measurements. The measurements of the magnetic GP are more challenging than those of the electric GP, but considering the experimental level of precision, it is expected that within this decade, we will be able to meaningfully disentangle the magnitude of the paramagnetic and diamagnetic contributions in the proton. The experimental measurements have provided high-precision benchmark data for the theory, and now, progress in the ChPT and LQCD calculations is needed in order to achieve a deep understanding of the key dynamical mechanisms that contribute to the electric and magnetic polarizability effects.

# Author contributions

NS: writing–original draft and writing–review and editing.

# Funding

The author(s) declare that financial support was received for the research, authorship, and/or publication of this article. This work was supported by the US Department of Energy Office of Science, Office of Nuclear Physics under contract no. DE-SC0016577.

# Acknowledgments

The author thank F. Hagelstein, J. Bericic, V. Pascalutsa, and F. Castellani for providing figures for this article.

# Conflict of interest

The author declares that the research was conducted in the absence of any commercial or financial relationships that could be construed as a potential conflict of interest.

The author(s) declared that they were an editorial board member of Frontiers, at the time of submission. This had no impact on the peer review process and the final decision.

## Publisher's note

All claims expressed in this article are solely those of the authors and do not necessarily represent those of their affiliated organizations, or those of the publisher, the editors, and the reviewers. Any product that may be evaluated in this article, or claim that may be made by its manufacturer, is not guaranteed or endorsed by the publisher.